\documentclass[floatfix,twocolumn,showpacs,prl,amsmath]{revtex4}
\usepackage{graphicx}
\usepackage{bm}
\usepackage{amssymb}
\usepackage{dcolumn}
\usepackage{subfigure}
\usepackage{threeparttable}
\usepackage{multirow}
\usepackage{mathrsfs}
\DeclareMathAlphabet{\mathscrbf}{OMS}{mdugm}{b}{n}
\usepackage[usenames,dvipsnames]{color}
\usepackage{mathtools}

\begin{document}
\newcommand{\vn}[1]{{\boldsymbol{#1}}}
\newcommand{\vht}[1]{{\boldsymbol{#1}}}
\newcommand{\matn}[1]{{\bf{#1}}}
\newcommand{\matnht}[1]{{\boldsymbol{#1}}}
\newcommand{\bege}{\begin{equation}}
\newcommand{\ee}{\end{equation}}
\newcommand{\bal}{\begin{aligned}}
\newcommand{\defbar}{\overline}
\newcommand{\SM}{\scriptstyle}
\newcommand{\eal}{\end{aligned}}
\newcommand{\torkance}{t}
\newcommand{\udot}{\overset{.}{u}}
\newcommand{\exponential}[1]{{\exp(#1)}}
\newcommand{\phandot}[1]{\overset{\phantom{.}}{#1}}
\newcommand{\phandag}{\phantom{\dagger}}
\newcommand{\Trace}{\text{Tr}}
\newcommand{\Bxc}{\Omega}
\newcommand{\mubo}{\mu_{\rm B}^{\phantom{B}}}
\newcommand{\rmd}{{\rm d}}
\newcommand{\magdir}{\hat{\vn{n}}}
\newcommand{\rme}{{\rm e}}
\newcommand{\intkspa}{\int\!\!\frac{\rmd^d k}{(2\pi)^d}}
\newcommand{\gret}{G_{\vn{k} }^{\rm R}(\mathcal{E})}
\setcounter{secnumdepth}{2}
\title{Geometrical contributions to the exchange constants: Free electrons with spin-orbit interaction}
\author{Frank Freimuth}
\email[Corresp.~author:~]{f.freimuth@fz-juelich.de}
\author{Stefan Bl\"ugel}
\author{Yuriy Mokrousov}
\affiliation{Peter Gr\"unberg Institut and Institute for Advanced Simulation,
Forschungszentrum J\"ulich and JARA, 52425 J\"ulich, Germany}
\date{\today}
\begin{abstract}
Using thermal quantum field theory we derive an expression for the
exchange constant that resembles Fukuyama's formula
for the orbital magnetic susceptibility (OMS). Guided by this formal analogy
between the exchange constant and OMS
we identify a contribution to the exchange constant that arises from the
geometrical properties of the band structure in mixed phase space.
We compute the exchange constants for free electrons
and show that the geometrical contribution is generally important.
Our formalism allows us to study the exchange
constants in the presence of spin-orbit interaction (SOI). Thereby, we find
sizable differences between the exchange constants of
helical and cycloidal spin spirals.
Furthermore, we discuss how to calculate the exchange constants based
on a gauge-field approach 
in the case of the Rashba model with an additional exchange splitting
and show that the exchange constants obtained from this
gauge-field approach are in perfect agreement
with those obtained from the quantum field theoretical method. 
\end{abstract}


\maketitle
\section{Introduction}
While the Berry phase  
has been shown to be important for 
spin-dynamics~\cite{spin_wave_dynamics_real_crystals,adiabatic_dynamics_local_spin_moments,spin_dynamics_tddft}, 
less attention has been paid to geometrical aspects in the exchange constants.
Recently, it has been shown that the
Dzyaloshinskii-Moriya interaction (DMI), i.e., the asymmetric exchange, can be
computed from a Berry phase approach, in which
the geometrical properties of the electronic structure
in mixed phase space play a key 
role~\cite{mothedmisot,phase_space_berry,itsot,spicudmi}.
DMI describes the linear change of the free energy with 
gradients in the magnetization direction.  
The effect of such noncollinear magnetic textures
on conduction electrons can be accounted for by effective 
magnetic potentials~\cite{bruno_the_2004,gauge_fields_spintronics}. Since orbital
magnetism leads to a linear change of the free energy when an
external magnetic field is applied,  
several formal analogies exist between the
modern theory of orbital magnetization~\cite{resta_review_om} 
and the Berry-phase 
approach to DMI~\cite{mothedmisot,itsot}, because the latter
captures the free energy change linear in an effective magnetic 
potential generated by the noncollinear magnetic texture. 

Similarly, the (symmetric) exchange constants describe the
quadratic change of the free energy with gradients in the
magnetization direction while the orbital magnetic susceptibility (OMS)
captures the quadratic change of the free energy with an applied
magnetic field~\cite{oms_fukuyama}. Therefore, it is
natural to suspect 
formal analogies between the
theories of OMS on the one hand and 
exchange constants on the other hand, which we will
investigate in detail in this paper.
For this purpose, we use thermal quantum field theory in
order to express the exchange constants in terms of
torque operators, velocity operators and the Green's functions
of a collinear ferromagnet and obtain a formula that resembles
Fukuyama's result for OMS~\cite{oms_fukuyama,oms_ogata_fukuyama}.

Recently, geometrical contributions to OMS have been identified
and shown to be generally significant and sometimes even
dominant~\cite{geometrical_effects_oms,quantum_metric_without_berry}.
These contributions arise from the 
reciprocal-space Berry curvature and
quantum metric, which describe geometrical properties of the
electronic structure. We will show that, 
as a consequence of the formal analogies
between OMS and exchange, similar geometrical
contributions to the exchange constants can be identified, which
arise from the Berry curvature and the quantum metric in mixed
phase space as well as from the quantum metric
in real space.
In order to achieve this, we rewrite our Fukuyama-type formula
for the exchange constant in terms of these geometrical properties.

Both the Fukuyama-type formula as well as the geometrical expression
allow us to obtain the exchange constants directly from the 
electronic structure. Compared to the frozen spin-spiral 
approach~\cite{adiabatic_spin_dynamics_dft_fe_co_ni,ab_initio_noco_flapw} 
such a formulation has the advantage 
that it becomes easier to investigate the relationship to 
spintronic and spincaloritronic effects. For example, the Berry phase theory
of DMI allows us to relate DMI to the spin-orbit torque~\cite{mothedmisot}, 
to ground-state 
spin-currents~\cite{spicudmi}, and to ground-state energy
currents which need to be subtracted in order to extract the
inverse thermal spin-orbit torque~\cite{itsot}.
Similarly, torques due to the exchange interaction need to be
considered in the theory of thermally induced spin-transfer 
torques~\cite{thermal_stt}, and a Green's function expression
of exchange is well suited for this purpose. 

For the calculation of exchange constants
in realistic materials powerful techniques
exist already. Besides the frozen
spin-spiral approach~\cite{adiabatic_spin_dynamics_dft_fe_co_ni,ab_initio_noco_flapw,
exchange_heuslers,
exchange_interactions_noco_dft} the
method of infinitesimal rotations of magnetic moments
and the Lichtenstein formula are 
popular~\cite{force_theorem,
exchange_dms,anisotropic_exchange_coupling_dms_dft}.
In this work we focus on free electrons. However, the 
extension of the Fukuyama-type approach
to calculations of exchange constants in
realistic materials within the framework of
first-principles density-functional theory has 
promising practical and technical perspectives.
For example, a Fukuyama-type formula
for the exchange constants might be an attractive alternative
when spin-orbit interaction (SOI) is 
present, because in this case the frozen spin-spiral approach 
cannot be used
and one needs to resort to
supercell methods or use multiple scattering 
theory~\cite{anisotropic_exchange_coupling_dms_dft}, which cannot be
combined easily with all available density-functional theory codes. 
Similarly, for the first-principles simulation of the current-induced motion of
domain walls and skyrmions, which involves complicated effects such as
chiral 
damping~\cite{chiral_damping_magnetic_domain_walls,phenomenology_chiral_damping} 
and the nonadiabatic 
torque~\cite{first_principles_nonadiabatic_stt}, 
and for the calculation of electronic
transport properties -- such as the topological Hall effect~\cite{the_mnsi} --
in these noncollinear magnetic textures an approach
that specifies the response to applied electric currents in terms of
a coefficient matrix that is expanded in orders of the magnetization gradients
is desirable. Since exchange constants are well-known for many materials,
their calculation from a Fukuyama-type expression can be used for code-testing
with the goal to extend the method to the mentioned spintronics effects.

This paper is structured as follows:
In section~\ref{sec_oms_fukuyama} we
briefly review the derivation of Fukuyama's
formula for OMS, which
serves as a basis to derive a Fukuyama-type
expression for the exchange constants in 
section~\ref{sec_exchange_fukuyama}.
In section~\ref{sec_oms_semicla} we discuss how to
express OMS in terms of reciprocal-space curvatures and
quantum metrices, which sets the stage to 
express the exchange constants in terms of
mixed phase space curvatures and quantum metrices
in section~\ref{sec_exchange_semicla}.
In section~\ref{sec_gauge_field} we show 
that -- despite the spin-orbit interaction -- the
exchange constants can be obtained easily from
a gauge-field approach in the case of the one-dimensional
Rashba model. In section~\ref{sec_one_dim_rashba}
we discuss the exchange constants of the
one-dimensional Rashba model. We show that the results obtained from the
Fukuyama-type approach agree to those of the gauge-field approach,
thereby demonstrating the validity of the Fukuyama-type expression
even in the presence of SOI. Additionally, we discuss the geometrical
contributions.
In section~\ref{sec_two_dim_rashba} we investigate the 
exchange constants in the two-dimensional Rashba model.
This paper ends with a summary in section~\ref{sec_summary}.
\section{Fukuyama method}
\subsection{Orbital magnetic susceptibility}
\label{sec_oms_fukuyama}
The orbital magnetic susceptibility tensor $\vn{\chi}$ is defined by
\bege\label{eq_def_oms}
\delta \vn{M}_{\rm orb}=\frac{1}{\mu_{0}}\vn{\chi} \vn{B},
\ee
where $\vn{B}$ is an applied external magnetic field and
$\delta \vn{M}_{\rm orb}$ is the change of the orbital magnetization due to 
the application of $\vn{B}$. $\mu_{0}$ is the vacuum permeability.
The $zz$ element of the orbital magnetic susceptibility tensor
is given by the Fukuyama formula~\cite{oms_fukuyama}
\bege\label{eq_oms_fukuyama}
\begin{aligned}
&\chi^{zz}=\frac{\mu_{0}e^2}{2\beta\hbar^2}
\intkspa
\sum_{p}
\Trace
\Bigl[\\
&G^{\rm M}_{\vn{k}}(i\mathcal{E}_{p})
v^{x}_{\vn{k}}
G^{\rm M}_{\vn{k}}(i\mathcal{E}_{p})
v^{y}_{\vn{k}}
G^{\rm M}_{\vn{k}}(i\mathcal{E}_{p})
v^{x}_{\vn{k}}
G^{\rm M}_{\vn{k}}(i\mathcal{E}_{p})
v^{y}_{\vn{k}}
\Bigr],
\end{aligned}
\ee
where $d$ is the dimension ($d$=2 or $d$=3). In the case of
twodimensional
systems, such as a graphene sheet or a thin film, the $z$ direction is
oriented perpendicular to the sheet or thin 
film. $v^{x}_{\vn{k}}$ and  $v^{y}_{\vn{k}}$ are the $x$ and $y$ components of the
velocity operator $\vn{v}_{\vn{k}}^{\phantom{k}}=e^{-i\vn{k}\cdot\vn{r}}\vn{v}e^{i\vn{k}\cdot\vn{r}}$ in 
crystal momentum representation, 
respectively. $\beta=(k_{\rm B}T)^{-1}$ is the
inverse temperature, $k_{\rm B}$ is the Boltzmann constant, 
and $\mathcal{E}_{p}=\beta^{-1}(2p+1)\pi$ are the Matsubara points.
\bege
G^{\rm M}_{\vn{k}}(i\mathcal{E}_{p})=
\hbar
[
i\mathcal{E}_{p}-H_{\vn{k}}
]^{-1}
\ee
is the Matsubara Green's function, 
where $H_{\vn{k}}$ is the Hamiltonian in crystal momentum representation.

Using the residue theorem the summation over Matsubara points
can be replaced by an energy integration along the real energy axis as follows:
\bege\label{eq_oms_fukuyama_real_axis}
\begin{aligned}
&\chi^{zz}
=-\frac{\mu_{0}e^2}{2\pi \hbar^2}
\intkspa
{\rm Im}\int d\,\mathcal{E}\,f(\mathcal{E})
\Trace
\Bigl[\\
&G^{\rm R}_{\vn{k}}(\mathcal{E})
v^{x}_{\vn{k}}
G^{\rm R}_{\vn{k}}(\mathcal{E})
v^{y}_{\vn{k}}
G^{\rm R}_{\vn{k}}(\mathcal{E})
v^{x}_{\vn{k}}
G^{\rm R}_{\vn{k}}(\mathcal{E})
v^{y}_{\vn{k}}
\Bigr],
\end{aligned}
\ee
where $f(\mathcal{E})$ is the Fermi function and
\bege
G^{\rm R}_{\vn{k}}(\mathcal{E})=\hbar
[\mathcal{E}-H_{\vn{k}}+i0^{+}]^{-1}
\ee 
is the retarded Green's function.

In the following we briefly sketch Fukuyama's 
derivation~\cite{oms_fukuyama} of 
Eq.~\eqref{eq_oms_fukuyama}, which serves as a preparation 
for obtaining an expression for the exchange constants
in section~\ref{sec_exchange_fukuyama}.
Since the vector potential of a homogeneous magnetic field
is not compatible with Bloch boundary conditions we
consider the spatially oscillating vector potential
\bege
\vn{A}(x)=\frac{B_{0}}{q}\sin(q x)\hat{\vn{e}}_{y}
\ee
with corresponding magnetic field
\bege
\vn{B}(x)=\nabla\times\vn{A}(x)=B_{0}\cos(qx)\hat{\vn{e}}_{z},
\ee
where $\hat{\vn{e}}_{y}$ and $\hat{\vn{e}}_{z}$ are unit vectors in 
the $y$ and $z$ directions, respectively.
At the final stage of the calculation the limit $q\rightarrow 0$ will be taken.
According to Eq.~\eqref{eq_def_oms} this spatially oscillating
magnetic field induces a spatially oscillating orbital magnetization.
The interaction between this induced orbital magnetization and
the magnetic field modifies the free energy density by the amount
\bege\label{eq_free_energy_oms}
\begin{aligned}
\delta F&=-\frac{1}{2}
\langle
\delta M_{\rm orb}^{z}
B^{z}_{\phantom{z}}
\rangle=
-\frac{1}{2\mu_{0}}
\chi^{zz}
\langle
B_{\phantom{z}}^{z}
B^{z}_{\phantom{z}}
\rangle=\\
&=-\frac{1}{4\mu_{0}}
\chi^{zz}
[B_{0}]^2,
\end{aligned}
\ee
where $\langle\dots\rangle$ denotes spatial averaging.
The expression for $\chi^{zz}$ can be 
found by determining $\delta F$ from thermal quantum field theory
and equating the result with 
Eq.~\eqref{eq_free_energy_oms}.

The free energy is obtained from the partition function $\Xi$
as
\bege
F=-\frac{1}{\beta}\ln\Xi.
\ee
The modification of $\Xi$ due to the applied magnetic field $\vn{B}(x)$
is determined from perturbation theory. For example the 
contribution from second order perturbation theory
is given by
\bege
\begin{aligned}
\Xi^{(2)}=&\frac{1}{2\hbar^2}
\int_{0}^{\hbar\beta}\!\!\!\! d\tau_{1}
\int_{0}^{\hbar\beta}\!\!\!\! d\tau_{2}
\,\Trace
\left[
e^{-\beta H}
T_{\tau}
\delta H_{\rm I}(\tau_{1})
\delta H_{\rm I}(\tau_{2})
\right]\\
=&\frac{\Xi^{(0)}}{2\hbar^2}
\int_{0}^{\hbar\beta}\!\!\!\! d\tau_{1}
\int_{0}^{\hbar\beta}\!\!\!\! d\tau_{2}
\,
\langle
T_{\tau}
\delta H_{\rm I}(\tau_{1})
\delta H_{\rm I}(\tau_{2})
\rangle,
\end{aligned}
\ee
where $\Xi^{(0)}$ is the partition function of the unperturbed system,
$T_{\tau}$ is the time-ordering operator, $H$ is the unperturbed Hamiltonian,
and $\delta H_{\rm I}(\tau)=e^{\tau H/\hbar}\delta H e^{-\tau H/\hbar}$ denotes the 
perturbation in the interaction picture.

Minimal coupling leads to two perturbation terms,
\bege\label{eq_minicoup1}
\delta H^{(1)}=\frac{e}{2}
\left[
\vn{v}\cdot\vn{A}(x)
+
\vn{A}(x) \cdot \vn{v}
\right]
\ee
and
\bege\label{eq_minicoup2}
\begin{aligned}
\delta H^{(2)}&=\frac{e^2}{2m_e}\vn{A}^2(x)=
\frac{e^2B_{0}^2}{2m_e q^2}\sin^2(qx)
=\\
&=\frac{e^2B_{0}^2}{4m_e q^2}
[1-\cos(2qx)],
\end{aligned}
\ee
where $e>0$ is the elementary positive 
charge and $m_e$ is the electron mass.
In order to determine $\chi^{zz}$ 
from Eq.~\eqref{eq_free_energy_oms}
we need to find the modification of the free energy density $\delta F$
that arises from the perturbations $\delta H^{(1)}$ and $\delta H^{(2)}$
and that is second order in $B_{0}$.
Thus, we need to perform second order perturbation theory with $\delta H^{(1)}$
and first order perturbation theory with $\delta H^{(2)}$.

In second quantization the perturbation $\delta H^{(1)}$ is given by
\bege
\begin{aligned}
\delta H^{(1)} &=
\frac{e
B_{0}
}
{4iq}
\sum_{\vn{k}nm}
\Bigl\{\\
&
\bigl[
\langle
u_{\vn{k}_{+}n}^{\phantom{y}}|
v_{\vn{k}_{+}}^{y}|
u_{\vn{k}_{-}m}^{\phantom{y}}
\rangle
+
\langle
u_{\vn{k}_{+}n}^{\phantom{y}}|
v_{\vn{k}_{-}}^{y}|
u_{\vn{k}_{-}m}^{\phantom{y}}
\rangle
\bigr]
c^{\dagger}_{\vn{k}_{+}n}c^{\phantom{\dagger}}_{\vn{k}_{-}m}\\
-&
\bigl[
\langle
u_{\vn{k}_{-}n}^{\phantom{y}}|
v_{\vn{k}_{-}}^{y}|
u_{\vn{k}_{+}m}^{\phantom{y}}
\rangle
+
\langle
u_{\vn{k}_{-}n}^{\phantom{y}}|
v_{\vn{k}_{+}}^{y}|
u_{\vn{k}_{+}m}^{\phantom{y}}
\rangle
\bigr]
c^{\dagger}_{\vn{k}_{-}n}c^{\phantom{\dagger}}_{\vn{k}_{+}m}
\Bigr\},
\end{aligned}
\ee
where $\vn{k}_{+}=\vn{k}+\vn{q}/2$ and $\vn{k}_{-}=\vn{k}-\vn{q}/2$
and $\vn{q}=q\hat{\vn{e}}_{x}$.
$|u_{\vn{k}n}\rangle$ denotes the eigenfunctions of the unperturbed 
Hamiltonian $H_{\vn{k}}$, such 
that $H_{\vn{k}}|u_{\vn{k}n}\rangle=\mathcal{E}_{\vn{k}n}|u_{\vn{k}n}\rangle$, 
where $\mathcal{E}_{\vn{k}n}$ is the band 
energy. $c^{\dagger}_{\vn{k}n}$ 
and $c^{\phantom{\dagger}}_{\vn{k}n}$ 
are creation and
annihilation operators of an electron in band $n$ at $k$-point $\vn{k}$, respectively.
Second order perturbation theory with respect to $\delta H^{(1)}$ modifies
the free energy density by the amount
\bege\label{eq_delta_frenergy}
\delta F\!=\!\frac{e^2 B^2_{0}}{4 q^2\beta\hbar^2}
\!\!\!\intkspa\!
\sum_{p}
\!\Trace\!
\left[
G^{\rm M}_{\vn{k}_{+}}
(i\mathcal{E}_{p})
v^{y}_{\vn{k}}
G^{\rm M}_{\vn{k}_{-}}(i\mathcal{E}_{p})
v^{y}_{\vn{k}}
\right].
\ee
When the trace in Eq.~\eqref{eq_delta_frenergy}
is Taylor-expanded in $q$ the zeroth-order term 
leads to a contribution to $\delta F$ that 
diverges like $q^{-2}$ in the limit $q\rightarrow 0$.
This divergent term cancels out with the 
contribution from the piece $e^2B_{0}^2/(4m_e q^2)$ in $\delta H^{(2)}$.
The oscillating piece $-e^2B_{0}^2\cos(2qx)/(4m_e q^2)$ in $\delta H^{(2)}$ averages
out in first order perturbation theory. 
The $q$-quadratic term from the Taylor-expansion of the trace in Eq.~\eqref{eq_delta_frenergy}
yields the free-energy change
\bege
\delta F\!=\!-\frac{e^2 B_{0}^2}{8\beta\hbar^2}
\!\!\!\intkspa\!
\sum_{p}
\Trace
\left[
\frac{\partial
G^{\rm M}_{\vn{k}}(i\mathcal{E}_{p})
}
{
\partial k^{x}_{\phantom{x}}
}
v^{y}_{\vn{k}}
\frac{
\partial G^{\rm M}_{\vn{k}}(i\mathcal{E}_{p})
}
{
\partial k^{x}_{\phantom{x}}
}
v^{y}_{\vn{k}}
\right].
\ee
With the help of Eq.~\eqref{eq_free_energy_oms}
we obtain the susceptibility
\bege
\chi^{zz}\!=\!\frac{e^2 \mu_{0}}{2\beta\hbar^2}
\!\!\!\intkspa\!
\sum_{p}
\Trace
\left[
\frac{\partial
G^{\rm M}_{\vn{k}}(i\mathcal{E}_{p})
}
{
\partial k^{x}_{\phantom{x}}
}
v^{y}_{\vn{k}}
\frac{
\partial G^{\rm M}_{\vn{k}}(i\mathcal{E}_{p})
}
{
\partial k^{x}_{\phantom{x}}
}
v^{y}_{\vn{k}}
\right].
\ee
Employing the relation
\bege\label{eq_k_deriv_green}
\frac{\partial G^{\rm M}_{\vn{k}}(i\mathcal{E}_{p})}{\partial k^{x}}=
G^{\rm M}_{\vn{k}}(i\mathcal{E}_{p})
v^{x}_{\vn{k}}
G^{\rm M}_{\vn{k}}(i\mathcal{E}_{p})
\ee
one finally obtains Eq.~\eqref{eq_oms_fukuyama}.

For completeness, we mention that it has been 
shown that Eq.~\eqref{eq_oms_fukuyama} needs to be modified
for the calculation of OMS from tight-binding 
models~\cite{orbital_magnetism_coupled_bands,lattice_effects_response_graphene}.
We do not discuss these modifications here.

\subsection{Exchange constants}
\label{sec_exchange_fukuyama}
In order to derive an expression for the 
exchange constant 
we consider the case where the magnetization 
performs small sinusoidal oscillations around the
$z$ direction as a function of the $x$ coordinate: 
\bege\label{eq_oszi_xz}
\hat{\vn{n}}(x)
=\begin{pmatrix}
\eta\sin(qx)\\
0\\
1
\end{pmatrix}
\frac{1}{\sqrt{1+\eta^2\sin^2(qx)}},
\ee
where $\hat{\vn{n}}(x)$ is a normalized vector
that describes the magnetization direction and
$\eta$ controls the amplitude of the oscillations.
As a result of these oscillations
the free energy density changes by the amount
\bege\label{eq_df_axx}
\delta F=\mathscr{A}^{xx}
\left\langle
\left[
\frac{
\partial
\hat{n}^{x}
}
{
\partial x
}
\right]^2
\right\rangle
=
\frac{1}{2}\eta^2q^2\mathscr{A}^{xx}
,
\ee
where $\mathscr{A}^{xx}$ is an exchange constant and
where we neglected higher orders in $\eta$.
In the presence of SOI the free energy change may depend
on whether the magnetization oscillates in the $xz$ plane or
in the $yz$ plane. When the magnetization oscillates
in the $yz$ plane, i.e., when 
\bege
\hat{\vn{n}}(x)
=\begin{pmatrix}
0\\
\eta\sin(qx)\\
1
\end{pmatrix}
\frac{1}{\sqrt{1+\eta^2\sin^2(qx)}},
\ee
the corresponding free energy change is described by
\bege
\delta F=\mathscr{A}^{xy}
\left\langle
\left[
\frac{
\partial
\hat{n}^{y}
}
{
\partial x
}
\right]^2
\right\rangle
=\frac{1}{2}\eta^2 q^2 \mathscr{A}^{xy},
\ee
with the exchange constant $\mathscr{A}^{xy}$. $\mathscr{A}^{xy}$
may differ from $\mathscr{A}^{xx}$ in the presence of SOI.
In the following we use thermal quantum field theory in order to
obtain expressions for the free energy change $\delta F$ that
arises from spatial oscillations of the magnetization direction as
given by Eq.~\eqref{eq_oszi_xz}. We will then use 
Eq.~\eqref{eq_df_axx} to obtain $\mathscr{A}^{xx}$.
To simplify the notation we will focus on the component $\mathscr{A}^{xx}$.
The generalization to the other exchange constants, such as $\mathscr{A}^{xy}$,
is obvious.

We consider the Hamiltonian of a collinear ferromagnet with
magnetization pointing in $z$ direction, given by
\bege\label{eq_hamil_coll_ferro}
\begin{aligned}
H(\vn{r})=&-\frac{\hbar^2}{2m_e}\Delta+V(\vn{r})+
\mubo\sigma^{z}\Bxc^{\rm xc}(\vn{r})+\\
&+
\frac{1}{2 e c^2}\mubo
\vn{\sigma}\cdot
\left[
\vn{\nabla}V(\vn{r})\times\vn{v}
\right].
\end{aligned}
\ee
The kinetic energy is described by the first term. The second term is
a scalar potential. The third term describes the exchange interaction,
where $\mubo$ is the
Bohr magneton, $\vn{\sigma}=(\sigma^{x},\sigma^{y},\sigma^{z})^{\rm T}$ 
is the vector of Pauli spin matrices, 
and $\Bxc^{\rm xc}(\vn{r})$ is the exchange field.
The last term is the spin-orbit interaction.
When the magnetization direction is not collinear
but 
spatially oscillating according to Eq.~\eqref{eq_oszi_xz}
the corresponding Hamiltonian is $H'=H+\delta H^{(1)}+\delta H^{(2)}$ with
\bege\label{eq_delta_h1_mag}
\delta H^{(1)}=
\mubo
\sigma^{x}
\Bxc^{\rm xc}(\vn{r})
\eta
\sin(qx)=\mathcal{T}^{y}\eta\sin(qx)
\ee
and
\bege\label{eq_delta_h2_mag}
\begin{aligned}
\delta H^{(2)}&=
-\frac{1}{2}\mubo
\sigma^{z}
\Bxc^{\rm xc}(\vn{r})
\eta^2
\sin^2(qx)\\
&=
-\frac{1}{4}\mubo
\sigma^{z}
\Bxc^{\rm xc}(\vn{r})
\eta^2
\left[1-
\cos(2qx)
\right],\\
\end{aligned}
\ee
where $\vn{\mathcal{T}}=-\mubo\vn{\sigma}\times\hat{\vn{e}}^{z}\Bxc^{\rm xc}$ 
is the torque operator and $\mathcal{T}^{y}$ is its $y$ component. 
According to Eq.~\eqref{eq_df_axx} we need to find the
modification of the free energy that is quadratic in $\eta$. Therefore,
we need to perform second order perturbation theory 
with $\delta H^{(1)}$
and first order perturbation theory with $\delta H^{(2)}$.

The perturbation $\delta H^{(1)}$ can be written in second quantization
in the form
\bege
\begin{aligned}
\delta H^{(1)}=\frac{\eta}{2i}
\sum_{\vn{k}nm}\Bigl\{
&\langle
u_{\vn{k}_{+}n}^{\phantom{k}}|
\mathcal{T}^{y}|
u_{\vn{k}_{-}m}^{\phantom{k}}
\rangle
c^{\dagger}_{\vn{k}_{+}n}c^{\phantom{\dagger}}_{\vn{k}_{-}m}\\
-
&\langle
u_{\vn{k}_{-}n}^{\phantom{k}}|
\mathcal{T}^{y}|
u_{\vn{k}_{+}m}^{\phantom{k}}
\rangle
c^{\dagger}_{\vn{k}_{-}n}c^{\phantom{\dagger}}_{\vn{k}_{+}m}
\Bigr\}.
\end{aligned}
\ee
In second order perturbation theory with 
respect to $\delta H^{(1)}$ the free energy is
modified by the amount
\bege\label{eq_delta_frenergy_exi}
\delta F\!=\!\frac{\eta^2}{4 \beta\hbar^2}
\!\!\!\intkspa\!
\sum_{p}
\!\Trace\!
\left[
G^{\rm M}_{\vn{k}_{+}}
(i\mathcal{E}_{p})
\mathcal{T}^{y}
G^{\rm M}_{\vn{k}_{-}}(i\mathcal{E}_{p})
\mathcal{T}^{y}
\right].
\ee
The zeroth-order term in the Taylor expansion of $\delta F$
with respect to $q$ cancels out with the contribution from the
piece $-\frac{1}{4}\mubo\sigma^{z}\Bxc^{\rm xc}(\vn{r})\eta^2$
from $\delta H^{(2)}$ only when SOI is not included. This is
an interesting difference to the case of the orbital magnetic susceptibility
discussed below Eq.~\eqref{eq_delta_frenergy}, where the
corresponding cancellation happens always. This difference is
due to the fact that the magnetic anisotropy energy
gives rise to a contribution to $\delta F$ 
which in leading order is proportional to $\eta^2$
at the zeroth order in $q$.
The oscillating piece
$\frac{1}{4}\mubo
\sigma^{z}
\Bxc^{\rm xc}(\vn{r})
\eta^2
\cos(2qx)$
from $\delta H^{(2)}$
averages out in first order perturbation theory.
In order to obtain the exchange constant $\mathscr{A}^{xx}$
we need the $q$-quadratic term from the Taylor-expansion 
of $\delta F$, which is given by
\bege\label{eq_delta_frenergy_exi2}
\delta F\!=\!-\frac{\eta^2q^2}{8 \beta\hbar^2}
\!\!\!\intkspa\!
\sum_{p}
\!\Trace\!
\left[
\frac{
\partial
G^{\rm M}_{\vn{k}_{+}}
(i\mathcal{E}_{p})}
{\partial k^x}
\mathcal{T}^{y}
\frac{
\partial
G^{\rm M}_{\vn{k}_{-}}
(i\mathcal{E}_{p})}
{\partial k^x}
\mathcal{T}^{y}
\right].
\ee
Using Eq.~\eqref{eq_k_deriv_green} and
Eq.~\eqref{eq_df_axx} we find the
following
expression for the 
exchange constant:
\bege\label{eq_exchange_params_fukuyama}
\begin{aligned}
&\mathscr{A}^{xx}=\frac{-1}{4\beta\hbar^2}
\sum_{p}\intkspa
\Trace
\Bigl[\\
&G^{\rm M}_{\vn{k}}(i\mathcal{E}_{p})
\mathcal{T}^{y}
G^{\rm M}_{\vn{k}}(i\mathcal{E}_{p})
v^{x}_{\vn{k}}
G^{\rm M}_{\vn{k}}(i\mathcal{E}_{p})
\mathcal{T}^{y}
G^{\rm M}_{\vn{k}}(i\mathcal{E}_{p})
v^{x}_{\vn{k}}
\Bigr],
\end{aligned}
\ee
which strongly resembles the Fukuyama formula for 
OMS, Eq.~\eqref{eq_oms_fukuyama}. Apart from
the prefactor, Eq.~\eqref{eq_exchange_params_fukuyama}
differs from Eq.~\eqref{eq_oms_fukuyama} by the
replacement of the velocity operator $v^{y}_{\vn{k}}$ 
by the torque operator $\mathcal{T}^{y}$.

The summation over Matsubara points can be
expressed in terms of an energy integration along the
real energy axis yielding 
\bege\label{eq_exchange_params_real_axis}
\begin{aligned}
&\mathscr{A}^{xx}
=\frac{1}{4\pi \hbar^2}{\rm Im}\int d\,\mathcal{E}\,f(\mathcal{E})
\intkspa
\Trace
\Bigl[\\
&G^{\rm R}_{\vn{k}}(\mathcal{E})
\mathcal{T}^{y}
G^{\rm R}_{\vn{k}}(\mathcal{E})
v^{x}_{\vn{k}}
G^{\rm R}_{\vn{k}}(\mathcal{E})
\mathcal{T}^{y}
G^{\rm R}_{\vn{k}}(\mathcal{E})
v^{x}_{\vn{k}}
\Bigr].
\end{aligned}
\ee
The unit of the exchange constant as given by 
Eq.~\eqref{eq_exchange_params_fukuyama} 
or Eq.~\eqref{eq_exchange_params_real_axis} is
energy times length when $d=1$ and it is
energy when $d=2$ and it is
energy per length when $d=3$.
Consequently, the unit of the free energy density as given by Eq.~\eqref{eq_df_axx}
is energy per length when  $d=1$ and it is
energy per area when  $d=2$ and it is
energy per volume when  $d=3$.

We have mentioned in the previous section that the Fukuyama formula for OMS
needs to be modified for tight-binding 
models~\cite{orbital_magnetism_coupled_bands,lattice_effects_response_graphene}.
We expect similar modifications to be necessary when exchange constants
are computed from tight-binding models, but we leave 
the discussion of these modifications for future work. 

\section{Curvatures, quantum metrices, moments and polarizations}
\subsection{Orbital magnetic susceptibility}
\label{sec_oms_semicla}
As discussed 
by Ogata et al.\ 
in~\cite{oms_ogata_fukuyama} 
one can
express the velocity operators and Green's functions in 
Eq.~\eqref{eq_oms_fukuyama} in the representation of Bloch
eigenfunctions such that
\bege\label{eq_oms_fukuyama_eigenspace}
\begin{aligned}
&\chi^{zz}=\frac{\mu_{0} e^2\hbar^2}{2\beta}
\sum_{\substack{nn'\\l \,l'}}
\intkspa
\Bigl[
v^{x}_{\vn{k}nn'}
v^{y}_{\vn{k}n'l}
v^{x}_{\vn{k}ll'}
v^{y}_{\vn{k}l'n}
\Bigr]\\
&\times
\sum_{p}
\frac{1}{i\mathcal{E}_{p}-\mathcal{E}_{\vn{k}n}}
\frac{1}{i\mathcal{E}_{p}-\mathcal{E}_{\vn{k}n'}}
\frac{1}{i\mathcal{E}_{p}-\mathcal{E}_{\vn{k}l}}
\frac{1}{i\mathcal{E}_{p}-\mathcal{E}_{\vn{k}l'}}
,
\end{aligned}
\ee
where $\vn{v}^{\phantom{x}}_{\vn{k}nn'}=\langle u_{\vn{k}n}|\vn{v}_{\vn{k}}^{\phantom{x}}| u_{\vn{k}n'}\rangle$
denotes the matrix elements of the velocity 
operator, $\mathcal{E}_{\vn{k}n}$ is the energy of band $n$ at
$k$-point $\vn{k}$ and $| u_{\vn{k}n}\rangle$ is the corresponding
eigenstate 
of $H_{\vn{k}}$, i.e., $H_{\vn{k}}| u_{\vn{k}n}\rangle=\mathcal{E}_{\vn{k}n}| u_{\vn{k}n}\rangle$. 
The summations over Matsubara points can be carried out
with the help of partial fraction decomposition and with the identity
\bege\label{eq_matsubara_sum}
\frac{1}{\beta}\sum_{p}
\frac{1}{
[
i\mathcal{E}_{p}-\mathcal{E}_{\vn{k}n}
]^{m}
}=\frac{1}{(m-1)!}f_{\vn{k}n}^{(m-1)},
\ee
where $f_{\vn{k}n}^{(m-1)}$ is the $(m-1)$th derivative of the
Fermi function.
For example, when $n=n'=l=l'$ in 
Eq.~\eqref{eq_oms_fukuyama_eigenspace}
one uses Eq.~\eqref{eq_matsubara_sum}
with $m=4$, which leads to a contribution 
with the third derivative of the Fermi function.
In order to rewrite
high derivatives of the Fermi function 
in terms of lower derivatives one
employs 
integration by parts and
the relation
\bege\label{eq_kderiv_fermifunc}
\vn{v}_{\vn{k}n}^{\phantom{x}}
f^{(m+1)}_{\vn{k}n}=
\frac{1}{\hbar}\frac{\partial f^{(m)}_{\vn{k}n}}{\partial \vn{k}},
\ee 
where we 
defined $\vn{v}_{\vn{k}n}^{\phantom{x}}=\vn{v}_{\vn{k}nn}^{\phantom{x}}$.
Thereby one can achieve that only the
first derivative of the Fermi function occurs.
The resulting expression for
the orbital magnetic susceptibility 
can be written as
\bege\label{eq_oms_semicla}
\begin{aligned}
\chi^{zz}=&
\mu_{0}
\frac{e^2}{\hbar^2}\intkspa
\sum_{n}
\Biggl [
\frac{1}{12}f'_{\vn{k}n}
(
\alpha^{xx}_{\vn{k}n}
\alpha^{yy}_{\vn{k}n}
-
\alpha^{xy}_{\vn{k}n}
\alpha^{yx}_{\vn{k}n}
)
\\
-&f'_{\vn{k}n}
m^{z}_{\vn{k}n}
m^{z}_{\vn{k}n}
-\frac{\hbar^2}{4m_e}
f_{\vn{k}n}
(
g^{xx}_{\vn{k}n}
+
g^{yy}_{\vn{k}n}
)\\
+&\frac{3}{2}
f_{\vn{k}n}
\Omega^{z}_{\vn{k}n}
m^{z}_{\vn{k}n}\\
+&\frac{1}{4}
f_{\vn{k}n}
(
g^{xx}_{\vn{k}n}\alpha^{yy}_{\vn{k}n}
+
g^{yy}_{\vn{k}n}\alpha^{xx}_{\vn{k}n}
-2
g^{xy}_{\vn{k}n}\alpha^{yx}_{\vn{k}n}
)\\
+&
\frac{\hbar^2}{2}
f'_{\vn{k}n}
v^{x}_{\vn{k}n}
\frac{
\partial
\langle u_{\vn{k}n}|
}{\partial k^y}
[
v^{x}_{\vn{k}}
+v^{x}_{\vn{k}n}
]
\frac{\partial
|u_{\vn{k}n}\rangle
}{\partial k^y}\\
+&
\frac{\hbar^2}{2}
f'_{\vn{k}n}
v^{y}_{\vn{k}n}
\frac{
\partial
\langle u_{\vn{k}n}|
}{\partial k^x}
[
v^{y}_{\vn{k}}
+v^{y}_{\vn{k}n}
]
\frac{\partial
|u_{\vn{k}n}\rangle
}{\partial k^x}\\
-&
\frac{\hbar^2}{2}
f'_{\vn{k}n}
v^{x}_{\vn{k}n}
\frac{
\partial
\langle u_{\vn{k}n}|
}{\partial k^y}
[
v^{y}_{\vn{k}}
+v^{y}_{\vn{k}n}
]
\frac{\partial
|u_{\vn{k}n}\rangle
}{\partial k^x}\\
-&
\frac{\hbar^2}{2}
f'_{\vn{k}n}
v^{y}_{\vn{k}n}
\frac{
\partial
\langle u_{\vn{k}n}|
}{\partial k^x}
[
v^{x}_{\vn{k}}
+v^{x}_{\vn{k}n}
]
\frac{\partial
|u_{\vn{k}n}\rangle
}{\partial k^y}
\\
-&
2\hbar^2
f_{\vn{k}n}\sum_{m\ne n}
\frac{
\mathcal{M}^{z}_{\vn{k}mn}
\left[
\mathcal{M}^{z}_{\vn{k}mn}
\right]^{*}
}
{
\mathcal{E}_{\vn{k}n}-
\mathcal{E}_{\vn{k}m}
}
\Biggr ]
,
\end{aligned}
\ee
where
\bege
\alpha^{ij}_{\vn{k}n}=
\frac{\partial^2 \mathcal{E}_{\vn{k}n}}{\partial k^i \partial k^j}
\ee
is the $ij$ element of the inverse effective mass tensor,
\bege
m^{z}_{\vn{k}n}=-
{\rm Im}
\left[
\frac{
\partial
\langle u_{\vn{k}n}|
}{\partial k^x}
[
\mathcal{E}_{\vn{k}n}-H
]
\frac{\partial
|u_{\vn{k}n}\rangle
}{\partial k^y}
\right]
\ee
is the $z$ component of the orbital moment of the wavepacket associated with 
band $n$ at $k$-point $\vn{k}$~\cite{wave_packets_sundaram,berry_phase_correction_dos},
\bege
g^{ij}_{\vn{k}n}={\rm Re}
\left[
\frac{
\partial
\langle u_{\vn{k}n}|
}{\partial k^i}
\Bigl[
1-
|u_{\vn{k}n}\rangle
\langle u_{\vn{k}n}|
\Bigr]
\frac{\partial
|u_{\vn{k}n}\rangle
}{\partial k^j}
\right]
\ee
is the $ij$ element of the $\vn{k}$-space quantum metrical 
tensor~\cite{geometry_quantum_evolution,quantum_geometry_bloch_bands,quantum_metric_without_berry}, 
$m_e$ is the electron mass,
\bege
\Omega^{z}_{\vn{k}n}=-2{\rm Im}
\left[
\frac{
\partial
\langle u_{\vn{k}n}|
}{\partial k^x}
\frac{\partial
|u_{\vn{k}n}\rangle
}{\partial k^y}
\right]
\ee
is the $\vn{k}$-space Berry curvature,
and
\bege
\vn{\mathcal{M}}_{\vn{k}mn}=
\frac{1}{2}
\left[
\sum_{n'\ne n}
\vn{v}_{\vn{k}mn'}\times\vn{A}_{\vn{k}n'n}
+
\vn{v}_{\vn{k}n}\times\vn{A}_{\vn{k}mn}
\right]
\ee
are interband matrix elements of the magnetic 
dipole moment and of the position operator~\cite{geometrical_effects_oms}, 
where
\bege\label{eq_k_space_connection}
\vn{A}_{\vn{k}mn}=i
\langle u_{\vn{k}m}|
\frac{\partial
|u_{\vn{k}n}\rangle
}{\partial \vn{k}}
=
i\hbar
\frac{
\langle u_{\vn{k}m}|
\vn{v}
|u_{\vn{k}n}\rangle
}
{
\mathcal{E}_{\vn{k}n}
-
\mathcal{E}_{\vn{k}m}
}
\ee
is the interband Berry connection.

A detailed discussion of all terms in Eq.~\eqref{eq_oms_semicla}
has been given by Gao et al.\ in Ref.~\cite{geometrical_effects_oms}.
In the semiclassical derivation of Gao et al.\ the terms in the 
lines 5, 6, 7 and 8 in Eq.~\eqref{eq_oms_semicla}
are explained by the $k$-space polarization energy and are related to the 
quadrupole moment of the velocity operator with 
respect to wave packets~\cite{geometrical_effects_oms}.
However, the semiclassical derivation yields a different prefactor for these
polarization terms. 
Already Ogata et al.\ pointed out in Ref.~\cite{oms_ogata_fukuyama} that the expression given 
by Gao et al.\ in Ref.~\cite{geometrical_effects_oms}
differs from Eq.~\eqref{eq_oms_fukuyama}. However, Ogata et al.\ compared the
semiclassical expression to the Fukuyama formula only in the special case of
space-inversion symmetric systems when time-reversal symmetry is not broken.
We find that Eq.~\eqref{eq_oms_fukuyama} can generally be written in the
form of Eq.~\eqref{eq_oms_semicla}, i.e., Eq.~\eqref{eq_oms_semicla} yields
the correct orbital magnetic susceptibility even in the time-reversal broken case
and in systems lacking space inversion symmetry.

Only the last line in Eq.~\eqref{eq_oms_semicla} involves 
interband couplings explicitly, while the first 8 lines in Eq.~\eqref{eq_oms_semicla}
are formulated in terms of single-band properties. 
The Berry curvature and the quantum metric 
describe the geometrical properties of a single-band. 
In this sense, lines 3 and 4 in Eq.~\eqref{eq_oms_semicla} 
constitute the geometrical
contribution to the orbital magnetic 
susceptibility~\cite{geometrical_effects_oms}. 
In section~\ref{sec_exchange_semicla} we will identify analogous
geometrical contributions to the exchange constants.

\subsection{Exchange constants}
\label{sec_exchange_semicla}
As discussed in section~\ref{sec_oms_semicla} the Fukuyama 
formula for the orbital magnetic susceptibility, Eq.~\eqref{eq_oms_fukuyama},
can be expressed in terms of 
geometrical properties such as the $k$-space Berry curvature
and the quantum metric, and several other single-band properties, such as the
orbital magnetic moment and the $k$-space polarization.
The expression for the exchange constants, Eq.~\eqref{eq_exchange_params_fukuyama},
has the same structure as Eq.~\eqref{eq_oms_fukuyama} and can be obtained by replacing
two velocity operators in Eq.~\eqref{eq_oms_fukuyama} by torque operators. This formal
similarity suggests that Eq.~\eqref{eq_exchange_params_fukuyama} can be expressed in
terms of Berry curvatures and quantum metrices in mixed phase space. 
For this purpose we define the mixed Berry curvature~\cite{phase_space_berry}
\bege\label{eq_mixed_curvature}
\mathcal{B}^{ij}_{\vn{k}n}=
-2\,
{\rm Im}
\left\langle
\frac{\partial u_{\vn{k}n}}{ \partial\hat{n}^{i} }
\left|
\frac{\partial u_{\vn{k}n}}{\partial k^{j}}\right.
\right\rangle,
\ee
where $\vn{k}$-derivatives are mixed with $\hat{\vn{n}}$-derivatives.
Similarly, we define the mixed quantum metric
\bege
\mathcal{G}^{ij}_{\vn{k}n}={\rm Re}
\left[
\frac{
\partial
\langle u_{\vn{k}n}|
}
{\partial \hat{n}^i}
\Bigl[
1-
|u_{\vn{k}n}\rangle
\langle u_{\vn{k}n}|
\Bigr]
\frac{\partial
|u_{\vn{k}n}\rangle
}
{\partial k^j}
\right].
\ee
Additionally, we define the quantum metric in magnetization space
\bege
\tilde{g}^{ij}_{\vn{k}n}={\rm Re}
\left[
\frac{
\partial
\langle u_{\vn{k}n}|
}{\partial \hat{n}^i}
\Bigl[
1-
|u_{\vn{k}n}\rangle
\langle u_{\vn{k}n}|
\Bigr]
\frac{\partial
|u_{\vn{k}n}\rangle
}{\partial \hat{n}^j}
\right].
\ee
The twist-torque moment of wavepackets is
described by~\cite{mothedmisot}
\bege
\mathcal{A}^{ij}_{\vn{k}n}=
-
{\rm Im}
\left\langle
\frac{\partial u_{\vn{k}n}}{ \partial\hat{n}^{i} }
\right|
\!\Bigl[
\mathcal{E}_{\vn{k}n}-H_{\vn{k}}
\Bigr]
\!\left|
\frac{\partial u_{\vn{k}n}}{\partial k^{j}}
\right\rangle
,
\ee
and
\bege
\bar{A}^{j}_{\vn{k}mn}=i
\langle u_{\vn{k}m}|
\frac{\partial
|u_{\vn{k}n}\rangle
}{\partial \hat{n}^{j}}
\ee
is the interband Berry connection in magnetization space. 
The mixed phase-space analogue of the inverse effective mass tensor is
given by
\bege\label{eq_mass_mixed}
\bar{\alpha}^{ij}_{\vn{k}n}=
\frac{\partial^2 \mathcal{E}_{\vn{k}n}}{\partial k^i \partial \hat{n}^j}.
\ee
In Appendix~\ref{sec_appendix_torque} we explain how the derivatives
with respect to magnetization direction are related to matrix elements
of the torque operator.

In terms of the mixed phase-space quantities Eq.~\eqref{eq_mixed_curvature} 
through Eq.~\eqref{eq_mass_mixed}
the exchange constant can be written as
\bege\label{eq_exchange_params_semicla}
\begin{aligned}
\mathscr{A}^{xx}&=
\intkspa
\sum_{n}
\Biggl [
\frac{1}{24}
(
f''_{\vn{k}n}
\mathcal{T}^{y}_{\vn{k}n}
\mathcal{T}^{y}_{\vn{k}n}
\alpha^{xx}_{\vn{k}n}
+
f'_{\vn{k}n}
\bar{\alpha}^{xx}_{\vn{k}n}
\bar{\alpha}^{xx}_{\vn{k}n}
)
\\
&+\frac{1}{2}f'_{\vn{k}n}
\mathcal{A}^{xx}_{\vn{k}n}
\mathcal{A}^{xx}_{\vn{k}n}
+\frac{1}{3}f_{\vn{k}n}
\tilde{g}^{xx}_{\vn{k}n}
\frac{\hbar^2}{m_e}
\\
&-\frac{5}{6}f_{\vn{k}n}
\mathcal{A}^{xx}_{\vn{k}n}
\mathcal{B}^{xx}_{\vn{k}n}
\\
&-\frac{1}{6}f_{\vn{k}n}
\alpha^{xx}_{\vn{k}n}
\tilde{g}^{xx}_{\vn{k}n}
+\frac{1}{6}f_{\vn{k}n}
\bar{\alpha}^{xx}_{\vn{k}n}
\mathcal{G}^{xx}_{\vn{k}n}
\\
&+\mathscr{A}^{xx}_{\rm pol}
+
\mathscr{A}^{xx}_{\rm inter}
\Biggr],
\end{aligned}
\ee
with
\bege\label{eq_axx_pol}
\begin{aligned}
\mathscr{A}^{xx}_{\rm pol}&=
\intkspa\sum_{n}
\Biggl [\\
&-\frac{1}{6}
f'_{\vn{k}n}
\mathcal{T}^{y}_{\vn{k}n}
\left\langle
\frac{\partial u_{\vn{k}n}}{ \partial k^{x}   }
\right|
[
\mathcal{T}^{y}_{\phantom{k}}
+2
\mathcal{T}^{y}_{\vn{k}n}
]
\left|
\frac{\partial u_{\vn{k}n}}{\partial k^{x}}
\right\rangle\\
&
-\frac{1}{6}\hbar^2
f'_{\vn{k}n}
v^{x}_{\vn{k}n}
\left\langle
\frac{\partial u_{\vn{k}n}}{ \partial  \hat{n}^{x} }
\right|
[
v^{x}_{\vn{k}}
+
v^{x}_{\vn{k}n}
]
\left|
\frac{\partial u_{\vn{k}n}}{\partial \hat{n}^{x}}
\right\rangle\\
&
+\frac{1}{3}\hbar
f'_{\vn{k}n}
\mathcal{T}_{\vn{k}n}^{y}
\left\langle
\frac{\partial u_{\vn{k}n}}{ \partial\hat{n}^{x} }
\right|
[
2v^{x}_{\vn{k}}
+
v^{x}_{\vn{k}n}
]
\left|
\frac{\partial u_{\vn{k}n}}{\partial k^{x}}
\right\rangle\Biggr],
\end{aligned}
\ee
and
\bege\label{eq_axx_inter}
\begin{aligned}
\mathscr{A}^{xx}_{\rm inter}&=
\intkspa\sum_{n}
\Biggl [
\frac{\hbar^2}{3}
f_{\vn{k}n}
v^{x}_{\vn{k}n}
v^{x}_{\vn{k}n}
\sum_{m\ne n}
\frac{
\bar{A}^{x}_{\vn{k}mn}
[\bar{A}^{x}_{\vn{k}mn}]^{*}
}
{\mathcal{E}_{\vn{k}n}-\mathcal{E}_{\vn{k}m}}
\\
&-
\frac{2}{3}
\hbar
f_{\vn{k}n}
v^{x}_{\vn{k}n}
\mathcal{T}^{y}_{\vn{k}n}
\sum_{m\ne n}
\frac{
\bar{A}^{x}_{\vn{k}mn}
[A^{x}_{\vn{k}mn}]^{*}
}
{\mathcal{E}_{\vn{k}n}-\mathcal{E}_{\vn{k}m}}
\\
&
-\frac{\hbar}{3}
f_{\vn{k}n}
\sum_{m\ne n}
\frac{
\sum\limits_{q\ne n}
[
v^{x}_{\vn{k}mq}
\bar{A}^{x}_{\vn{k}qn}
]^{*}
\sum\limits_{r\ne n}
\mathcal{T}^{y}_{\vn{k}mr}
A^{x}_{\vn{k}rn}
}
{\mathcal{E}_{\vn{k}n}-\mathcal{E}_{\vn{k}m}}\\
&
+\frac{2}{3}\hbar^2
f_{\vn{k}n}
\sum_{m\ne n}
\frac{
\sum\limits_{q\ne n}
[
v^{x}_{\vn{k}mq}
\bar{A}^{x}_{\vn{k}qn}
]^{*}
\sum\limits_{r\ne n}
v^{x}_{\vn{k}mr}
\bar{A}^{x}_{\vn{k}rn}
}
{\mathcal{E}_{\vn{k}n}-\mathcal{E}_{\vn{k}m}}\\
&
-\frac{1}{3}\hbar
f_{\vn{k}n}
v^{x}_{\vn{k}n}
\sum_{m\ne n}
\frac{
[
\bar{A}^{x}_{\vn{k}mn}
]^{*}
\sum\limits_{r\ne n}
\mathcal{T}^{y}_{\vn{k}mr}
A^{x}_{\vn{k}rn}
}
{\mathcal{E}_{\vn{k}n}-\mathcal{E}_{\vn{k}m}}\\
&
-\frac{2}{3}\hbar
f_{\vn{k}n}
\mathcal{T}^{y}_{\vn{k}n}
\sum_{m\ne n}
\frac{
[
A^{x}_{\vn{k}mn}
]^{*}
\sum\limits_{r\ne n}
v^{x}_{\vn{k}mr}
\bar{A}^{x}_{\vn{k}rn}
}
{\mathcal{E}_{\vn{k}n}-\mathcal{E}_{\vn{k}m}}\\
&
+
f_{\vn{k}n}
\mathcal{T}^{y}_{\vn{k}n}
\sum_{m\ne n}
\frac{
[
A^{x}_{\vn{k}mn}
]^{*}
\sum\limits_{r\ne n}
\mathcal{T}^{y}_{\vn{k}mr}
A^{x}_{\vn{k}rn}
}
{\mathcal{E}_{\vn{k}n}-\mathcal{E}_{\vn{k}m}}
\Biggr],
\end{aligned}
\ee
where we 
defined $\mathcal{T}^{y}_{\vn{k}nn'}=\langle u_{\vn{k}n}|\mathcal{T}^{y}_{\phantom{k}}| u_{\vn{k}n'}\rangle$
and $\mathcal{T}^{y}_{\vn{k}n}=\mathcal{T}^{y}_{\vn{k}nn}$.

Eq.~\eqref{eq_exchange_params_semicla}
differs substantially in structure from
Eq.~\eqref{eq_oms_semicla},
while the corresponding
Fukuyama-type expressions,
Eq.~\eqref{eq_oms_fukuyama}
and Eq.~\eqref{eq_exchange_params_fukuyama},
are very similar structurally.
The structural differences between 
Eq.~\eqref{eq_exchange_params_semicla}
and
Eq.~\eqref{eq_oms_semicla}
arise, because there is no integration over the
magnetization direction, only a 
Brillouin zone integration, and therefore the identity
\bege\label{eq_magderiv_fermifunc}
\vn{\mathcal{T}}_{\vn{k}n}^{\phantom{y}}
f^{(m+1)}_{\vn{k}n}=
\hat{\vn{n}}\times
\frac{\partial f^{(m)}_{\vn{k}n}}{\partial \hat{\vn{n}}}
\ee
cannot be combined with integration by parts 
in order to rewrite high derivatives of the Fermi
function in terms of lower derivatives of the Fermi function
while
Eq.~\eqref{eq_kderiv_fermifunc} can be used for this purpose.
For example, the first line in Eq.~\eqref{eq_exchange_params_semicla}
is related formally to the Landau-Peierls susceptibility in the
first line of Eq.~\eqref{eq_oms_semicla}:
In the case of the orbital magnetic susceptibility the
torque operators in the first line of Eq.~\eqref{eq_exchange_params_semicla}  
turn into velocity operators and one can use integration by parts such that
\bege
\begin{aligned}
&\intkspa
f''_{\vn{k}n}
v^{y}_{\vn{k}n}
v^{y}_{\vn{k}n}
\alpha^{xx}_{\vn{k}n}=
\intkspa
\frac{1}{\hbar}
\frac{
\partial
f'_{\vn{k}n}
}
{
\partial k^y
}
v^{y}_{\vn{k}n}
\alpha^{xx}_{\vn{k}n}=\\
-&\intkspa
\frac{1}{\hbar}
f'_{\vn{k}n}
\frac{
\partial}
{\partial k^y}
[
v^{y}_{\vn{k}n}
\alpha^{xx}_{\vn{k}n}]=\\
-&\intkspa
\frac{1}{\hbar}
f'_{\vn{k}n}
[
\alpha^{yy}_{\vn{k}n}
\alpha^{xx}_{\vn{k}n}
+
v^{y}_{\vn{k}n}
\frac{
\partial}
{\partial k^y}
\alpha^{xx}_{\vn{k}n}],\\
\end{aligned}
\ee
which contains the 
term $f'_{\vn{k}n}\alpha^{yy}_{\vn{k}n}\alpha^{xx}_{\vn{k}n}$
found also in the first line of Eq.~\eqref{eq_oms_semicla}.

The lines 2, 3 and 4 in Eq.~\eqref{eq_exchange_params_semicla}
correspond to the lines 2, 3 and 4 in 
Eq.~\eqref{eq_oms_semicla}, where the twist torque 
moment $\mathcal{A}_{\vn{k}n}^{xx}$
replaces the orbital moment $m^{z}_{\vn{k}n}$,
the $k$-space quantum metric $g_{\vn{k}n}^{yy}$ is replaced by
the magnetization-space quantum metric $\tilde{g}_{\vn{k}n}^{xx}$,
the mixed Berry curvature replaces the $k$-space Berry curvature,
and the off-diagonal elements of the inverse effective 
mass, $\alpha_{\vn{k}n}^{yx}$, and of
the $k$-space quantum metric, $g_{\vn{k}n}^{xy}$,
are replaced by their mixed phase-space counterparts.

The contribution $\mathscr{A}_{\rm pol}^{xx}$ 
defined in Eq.~\eqref{eq_axx_pol}
corresponds to the lines 5, 6, 7 and 8 in Eq.~\eqref{eq_oms_semicla},
which describe the $k$-space polarization energy.
The contribution $\mathscr{A}_{\rm inter}^{xx}$ 
defined in Eq.~\eqref{eq_axx_inter}
corresponds to the last line in Eq.~\eqref{eq_oms_semicla} 
and is the only
term that contains interband couplings explicitly.

Several terms in Eq.~\eqref{eq_exchange_params_semicla} are zero
when SOI is not included in the Hamiltonian:
The mixed phase-space analogue of the inverse effective 
mass, $\bar{\alpha}_{\vn{k}n}^{ij}$, is zero without SOI, because the
band energy does not depend on the magnetization direction when 
SOI is absent. Additionally, $\mathcal{T}_{\vn{k}n}^{y}=0$,
 $\mathcal{A}^{ij}_{\vn{k}n}=0$, $\mathcal{B}^{ij}_{\vn{k}n}=0$ 
and $\mathcal{G}^{ij}_{\vn{k}n}=0$ 
in the
absence of SOI.
Thus, when SOI is absent the exchange constants are given by the
considerably simpler expression
\bege\label{eq_exchange_params_semicla_nosoi}
\begin{aligned}
\mathscr{A}^{xx}&=
\intkspa
\sum_{n}
\Biggl [
\frac{1}{3}f_{\vn{k}n}
\tilde{g}^{xx}_{\vn{k}n}
\frac{\hbar^2}{m_e}
\\
&-\frac{1}{6}f_{\vn{k}n}
\alpha^{xx}_{\vn{k}n}
\tilde{g}^{xx}_{\vn{k}n}
\\
&
-\frac{\hbar^2}{6}
f'_{\vn{k}n}
v_{\vn{k}n}^{x}
\left\langle
\frac{\partial u_{\vn{k}n}}{ \partial  \hat{n}^{x} }
\right|
[
v^{x}_{\vn{k}}
+
v^{x}_{\vn{k}n}
]
\left|
\frac{\partial u_{\vn{k}n}}{\partial \hat{n}^{x}}
\right\rangle\\
&+
\frac{\hbar^2}{3}
f_{\vn{k}n}
v^{x}_{\vn{k}n}
v^{x}_{\vn{k}n}
\sum_{m\ne n}
\frac{
\bar{A}^{x}_{\vn{k}mn}
[\bar{A}^{x}_{\vn{k}mn}]^{*}
}
{\mathcal{E}_{\vn{k}n}-\mathcal{E}_{\vn{k}m}}
\\
&
-\frac{\hbar}{3}
f_{\vn{k}n}
\sum_{m\ne n}
\frac{
\sum\limits_{q\ne n}
[
v^{x}_{\vn{k}mq}
\bar{A}^{x}_{\vn{k}qn}
]^{*}
\sum\limits_{r\ne n}
\mathcal{T}^{y}_{\vn{k}mr}
A^{x}_{\vn{k}rn}
}
{\mathcal{E}_{\vn{k}n}-\mathcal{E}_{\vn{k}m}}\\
&
+\frac{2}{3}
\hbar^2
f_{\vn{k}n}
\sum_{m\ne n}
\frac{
\sum\limits_{q\ne n}
[
v^{x}_{\vn{k}mq}
\bar{A}^{x}_{\vn{k}qn}
]^{*}
\sum\limits_{r\ne n}
v^{x}_{\vn{k}mr}
\bar{A}^{x}_{\vn{k}rn}
}
{\mathcal{E}_{\vn{k}n}-\mathcal{E}_{\vn{k}m}}\\
&
-\frac{1}{3}\hbar
f_{\vn{k}n}
v^{x}_{\vn{k}n}
\sum_{m\ne n}
\frac{
[
\bar{A}^{x}_{\vn{k}mn}
]^{*}
\sum\limits_{r\ne n}
\mathcal{T}^{y}_{\vn{k}mr}
A^{x}_{\vn{k}rn}
}
{\mathcal{E}_{\vn{k}n}-\mathcal{E}_{\vn{k}m}}
\Biggr].
\end{aligned}
\ee
In Appendix~\ref{sec_analytical} we discuss how to evaluate
Eq.~\eqref{eq_exchange_params_semicla_nosoi} analytically for simple model systems.

The lines 3 and 4 in Eq.~\eqref{eq_exchange_params_semicla} are
the geometrical contribution to the exchange constants.
It consists of three terms:
\bege\label{eq_geo1}
\mathscr{A}_{\rm geo1}^{xx}=
-\frac{5}{6}
\intkspa
\sum_{n}
f_{\vn{k}n}
\mathcal{A}^{xx}_{\vn{k}n}
\mathcal{B}^{xx}_{\vn{k}n}
\ee
and
\bege\label{eq_geo2}
\mathscr{A}_{\rm geo2}^{xx}=
-\frac{1}{6}
\intkspa
\sum_{n}
f_{\vn{k}n}
\alpha^{xx}_{\vn{k}n}
\tilde{g}^{xx}_{\vn{k}n}
\ee
and
\bege\label{eq_geo3}
\mathscr{A}_{\rm geo3}^{xx}=
\frac{1}{6}
\intkspa
\sum_{n}
f_{\vn{k}n}
\bar{\alpha}^{xx}_{\vn{k}n}
\mathcal{G}^{xx}_{\vn{k}n}.
\ee
$\mathcal{B}^{xx}_{\vn{k}n}$ 
and $\mathcal{G}^{xx}_{\vn{k}n}$
describe geometrical properties of the bands in mixed
phase space.
When SOI is not included in the 
Hamiltonian $\mathscr{A}_{\rm geo1}^{xx}$ 
and $\mathscr{A}_{\rm geo3}^{xx}$ 
are zero.
$\mathscr{A}_{\rm geo2}^{xx}$ is nonzero even in the
absence of SOI. It involves $\tilde{g}^{xx}_{\vn{k}n}$, which
describes the geometrical properties of the bands in real space.

According to Eq.~\eqref{eq_axx_pol} $\mathscr{A}^{xx}_{\rm pol}$ 
contains only terms with $f'_{\vn{k}n}$. 
The derivative of the Fermi function becomes large close to
the Fermi energy. In particular at zero temperature we have
$f'_{\vn{k}n}=-\delta(\mathcal{E}_{\rm F}-\mathcal{E}_{\vn{k}n})$.
Therefore only states close 
to the Fermi level contribute to $\mathscr{A}^{xx}_{\rm pol}$, i.e.,
$\mathscr{A}^{xx}_{\rm pol}$ is a Fermi surface term.
In contrast, $\mathscr{A}^{xx}_{\rm inter}$
(Eq.~\eqref{eq_axx_inter}) 
contains only terms
with $f_{\vn{k}n}$, i.e., all states below the Fermi energy
contribute to $\mathscr{A}^{xx}_{\rm inter}$. 
Hence, $\mathscr{A}^{xx}_{\rm inter}$
is a Fermi sea term. 
Eq.~\eqref{eq_exchange_params_semicla}
contains additional Fermi surface and Fermi sea terms.
The exchange constant in magnetic band insulators arises from the
Fermi sea terms, since the Fermi surface terms are zero in insulators. 

\section{Gauge-field approach}
\label{sec_gauge_field}
The appearance of gauge-fields and their
application in spintronics has been discussed in detail
in the review Ref.~\cite{gauge_fields_spintronics}. They can
occur in real-space, momentum-space and in time.
Here, we are interested in the Berry gauge field associated with
electron spins that adiabatically follow noncollinear magnetic textures.
This gauge field mimics the magnetic vector potential known from
electrodynamics. The curl of this effective magnetic vector potential
has similar consequences like a real magnetic field.
In particular it deflects electrons by an effective Lorentz force, which
leads to the topological Hall effect~\cite{bruno_the_2004}. 
The curl of the effective magnetic vector potential is nonzero when
the scalar spin chirality of the magnetic texture is nonzero, for example in
skyrmions. For the discussion of the exchange constants it is not necessary
to consider systems with nonzero scalar spin chirality. But even when the curl
of the effective magnetic vector potential is zero it does have consequences,
in particular it affects the energy of the eigenstates, as we will see below. 

In the case of the topological Hall effect the gauge-field approach has been
developed for systems without SOI~\cite{bruno_the_2004}.
In the general case it is difficult to apply the gauge-field approach to magnetic
systems with SOI. However, under certain conditions the exchange constants
can be obtained from a gauge-field approach even in
the presence of SOI. We demonstrate this in the following.
We will show that the exchange constants calculated
based on the
gauge-field approach agree to those given 
by Eq.~\eqref{eq_exchange_params_fukuyama}.
This will prove the accuracy of Eq.~\eqref{eq_exchange_params_fukuyama}.

We consider the Rashba model with an additional 
exchange splitting 
(see Ref.~\cite{rashba_review} for a recent review
on the Rashba model)
\bege\label{eq_rashba_model}
H=\frac{-\hbar^2}{2m_e}
\Delta-i
\alpha (\vn{\nabla}\times\hat{\vn{e}}_{z})\cdot\vn{\sigma}+
\frac{\Delta V}{2}
\vn{\sigma}
\cdot
\magdir_{\rm c}(\vn{r})
,
\ee
where the first, second and third terms on the right-hand side
describe the kinetic energy,
the Rashba spin-orbit coupling
and the exchange interaction, respectively.
We focus on the case of a flat cycloidal
spin-spiral, where the magnetization direction
$\magdir_{\rm c}(\vn{r})$ is given by
\bege\label{eq_spin_spiral_cycloid}
\magdir_{\rm c}(\vn{r})=
\begin{pmatrix}
\sin(qx)\\
0\\
\cos(qx)
\end{pmatrix}.
\ee

The exchange interaction describing the
noncollinear spin-spiral in Eq.~\eqref{eq_rashba_model}
can be transformed into an effective exchange
interaction of a collinear magnet with the help of the
unitary transformation
\bege\label{eq_gauge_trafo_matrix}
U(x)=
\left(
\begin{array}{cc}
\cos(\frac{qx}{2}) &-\sin(\frac{qx}{2})\\[6pt]
\sin(\frac{qx}{2}) &\cos(\frac{qx}{2})
\end{array}
\right)
\ee
such that~\cite{bruno_the_2004}
\bege
U^{\dagger}(x)
\frac{\Delta V}{2}
\vn{\sigma}
\cdot
\magdir_{\rm c}(\vn{r})U(x)=
\frac{\Delta V}{2}
\sigma_{z}.
\ee
The kinetic energy in Eq.~\eqref{eq_rashba_model}
transforms under this unitary transformation as follows~\cite{bruno_the_2004}:
\bege
\begin{aligned}
&-\frac{\hbar^2}{2m_e}
U^{\dagger}
\Delta
U
=\\
&=-\frac{\hbar^2}{2m_e}
U^{\dagger}
\frac{\partial}{\partial \vn{r}}\cdot
\left(
U\frac{\partial}{\partial \vn{r}}
+\frac{\partial U}{\partial \vn{r}}
\right)\\
&=-\frac{\hbar^2}{2m_e}
\left(
\Delta
+2
U^{\dagger}
\frac{\partial U}{\partial x}
\frac{\partial}{\partial x}
+
U^{\dagger}
\frac{\partial^2 U}{\partial x^2}
\right).
\end{aligned}
\ee
The derivatives of $U$ with respect to 
the $x$ coordinate are
\bege\label{eq_gauge_trafo_matrix_der1}
\frac{\partial U(x)}{\partial x}=\frac{q}{2}
\left(
\begin{array}{cc}
-\sin(\frac{qx}{2}) &-\cos(\frac{qx}{2})\\[6pt]
\cos(\frac{qx}{2}) &-\sin(\frac{qx}{2})
\end{array}
\right)
\ee
and
\bege\label{eq_gauge_trafo_matrix_der2}
\frac{\partial^2 U(x)}{\partial x^2}=
-\frac{q^2}{4}U(x)
\ee
and we have
\bege
[U(x)]^{\dagger}
\frac{\partial U(x)}{\partial x}=\frac{q}{2}\left(
\begin{array}{cc}
0 &-1\\
1 &0
\end{array}
\right)=\frac{q}{2i}\sigma_{y}
\ee
such that the kinetic energy transforms as
\bege
-\frac{\hbar^2}{2m_e}
U^{\dagger}
\Delta
U
=-\frac{\hbar^2}{2m_e}
\left(
\Delta
-iq\sigma_y\frac{\partial}{\partial x}
-\frac{q^2}{4}
\right).\\
\ee

Next, we need to find out how the
Rashba SOI
\bege
\frac{1}{i}\alpha\vn{\sigma}\cdot(\vn{\nabla}\times\hat{\vn{e}}_z)
=\frac{1}{i}\alpha
\left[
\sigma_{x}
\frac{\partial}{\partial y}
-\sigma_{y}
\frac{\partial}{\partial x}
\right]
\ee
transforms under $U$.
We have
\bege
\begin{aligned}
[U(x)]^{\dagger}
\sigma_{y}
\frac{\partial U(x)}{\partial x}=-i\frac{q}{2}
\end{aligned}
\ee
and
\bege
\begin{aligned}
[U(x)]^{\dagger}
\sigma_{y}
U(x)=\sigma_{y}
\end{aligned}
\ee
and thus
\bege
\begin{aligned}
&-[U(x)]^{\dagger}
\left[
\frac{\alpha}{i}
\sigma_{y}
\frac{\partial}{\partial x}
\right]
U(x)=\\
=&-[U(x)]^{\dagger}
\left[
\alpha
\sigma_{y}
\right]
U(x)
\frac{1}{i}
\frac{\partial }{\partial x}\\
&-[U(x)]^{\dagger}
\left[
\frac{\alpha}{i}
\sigma_{y}
\right]
\frac{\partial
U(x)
}{\partial x}
=\\
=&-
\alpha
\sigma_{y}
\frac{1}{i}
\frac{\partial }{\partial x}
+\frac{\alpha q}{2}.\\
\end{aligned}
\ee
However
\bege\label{eq_gauge_trafo_sigmax}
\begin{aligned}
&[U(x)]^{\dagger}
\sigma_{x}
U(x)
=\\
&=
\left(
\begin{matrix}
2\cos^2(\frac{qx}{2})-1 
&-2
\cos(\frac{qx}{2})
\sin(\frac{qx}{2})\\[6pt]
-2
\cos(\frac{qx}{2})
\sin(\frac{qx}{2}) &
-2\cos^2(\frac{qx}{2})+1
\end{matrix}
\right)
\end{aligned}
\ee
depends on the $x$ coordinate and consequently
the application of 
the $U$ transformation to Eq.~\eqref{eq_rashba_model} 
transforms
the $x$-dependence of the exchange interaction 
into an $x$-dependence of SOI and
no simplification is achieved by this transformation.

Therefore, we consider now the one-dimensional version of the Rashba model
with an additional exchange splitting
\bege\label{eq_rashba_model_onedim}
H=\frac{-\hbar^2}{2m_e}
\frac{\partial^2}{\partial x^2}
+i
\alpha 
\sigma_{y}\frac{\partial}{\partial x}
+
\frac{\Delta V}{2}
\vn{\sigma}
\cdot
\magdir_{\rm c}(\vn{r}).
\ee
The one-dimensional Rashba model can be used to describe spin-split bands
in one-dimensional atomic chains on 
surfaces~\cite{spin_split_bands_onedim_chain}.
Application of the $U$ transformation to Eq.~\eqref{eq_rashba_model_onedim}
yields
\bege\label{eq_onedim_rashba_model_gauge}
\begin{aligned}
\tilde{H}=[U(x)]^{\dagger}
H
U(x)
&=
-\frac{\hbar^2}{2m_e}
\left[
\frac{\partial^2}{\partial x^2}
-iq\sigma_y\frac{\partial}{\partial x}
-\frac{q^2}{4}
\right]+\\
&+
\frac{\Delta V}{2}
\sigma_{z}
-
\alpha
\sigma_{y}
\frac{1}{i}
\frac{\partial }{\partial x}
+\frac{\alpha q}{2}
\end{aligned}
\ee
and the corresponding crystal-field representation of the
Hamiltonian is given by
\bege\label{eq_onedim_rashba_gauge_crystal_field}
\begin{aligned}
\tilde{H}_{k_x}
&=
\frac{\hbar^2}{2m_e}
\left[
k_x^2
-q
k_{x}
\sigma_y
+\frac{q^2}{4}
\right]+\\
&+
\frac{\Delta V}{2}
\sigma_{z}
-
\alpha
k_{x}
\sigma_{y}
+\frac{\alpha q}{2}.
\end{aligned}
\ee
Since the $U$ transformation preserves the eigenvalues, the 
Hamiltonian $\tilde{H}_{k_x}$ has the same spectrum
as Eq.~\eqref{eq_rashba_model_onedim}. However, $\tilde{H}_{k_x}$ 
is position-independent and thus it is straightforward to determine
its eigenvalues, while the original Hamiltonian in Eq.~\eqref{eq_rashba_model_onedim}
is more difficult to deal with due to the position-dependence of the
exchange term for the cycloidal spin-spiral.

The reason why the $U$ transformation can be used to
simplify Eq.~\eqref{eq_rashba_model_onedim} into $\tilde{H}_{k_x}$ lies in the
spin-rotation symmetry of Eq.~\eqref{eq_rashba_model_onedim}: 
The Hamiltonian is invariant under the simultaneous rotation of the spin-operator
and the magnetization direction $\hat{\vn{n}}_{c}$ around the $y$ axis.
In contrast, the Hamiltonian of the two-dimensional 
Rashba model, Eq.~\eqref{eq_rashba_model},
does not exhibit this symmetry when $\alpha\neq 0$.

The Hamiltonian Eq.~\eqref{eq_onedim_rashba_model_gauge} can be rewritten in the form
\bege
H=\frac{1}{2m}
(p_x+eA^{\rm eff})^2-\frac{m \alpha^2}{2  \hbar^2}
\ee
where $p_x=-i\hbar \partial/\partial x$ is the $x$ component of the
momentum operator and
\bege
A^{\rm eff}=-\frac{m }{e  \hbar}
\left(
\alpha+\frac{\hbar^2}{2m}q
\right)
\sigma_{y}
\ee
can be considered as an effective magnetic vector potential, which is
why 
we refer to this method as gauge-field approach.

The free energy density $F_{q}$ of the one-dimensional
Rashba model with exchange splitting, Eq.~\eqref{eq_rashba_model_onedim},
can be obtained from 
\bege
F_{q}=
-\frac{1}{\beta}
\int\frac{d\,k_{x}}{2\pi}\sum_{n}
\ln
\left[
1+e^{-\beta(\mathcal{E}_{k_x,q,n}-\mu)}
\right]
,
\ee
where $\mathcal{E}_{k_x,q,n}$ denotes the $n$th eigenvalue of $\tilde{H}_{k_x}$
at $k$-point $k_{x}$ and spin-spiral wavenumber $q$.
Equating $F_{q}$  and the phenomenological expression for the free energy
\bege
F_{q}=F_{0}+D^{yx}q+\mathscr{A}^{xx}q^2
\ee
allows us to determine the DMI-coefficient and the exchange parameter
as follows:
\bege\label{eq_gauge_approach_dmi}
D^{yx}=\frac{F_{q}-F_{-q}}{2q}
\ee
and
\bege\label{eq_gauge_approach_exchange}
\mathscr{A}^{xx}=\frac{
F_{q}+F_{-q}
-2F_{0}
}{2 q^2}.
\ee
In section~\ref{sec_one_dim_rashba} we will compare the 
exchange constant $\mathscr{A}^{xx}$ obtained from 
Eq.~\eqref{eq_gauge_approach_exchange} to the one given
by Eq.~\eqref{eq_exchange_params_fukuyama} and find 
perfect agreement between these two rather different
approaches. Additionally, 
we will show in section~\ref{sec_one_dim_rashba} that
the
DMI-coefficient $D^{yx}$ obtained
from Eq.~\eqref{eq_gauge_approach_dmi}
is in perfect agreement
with the one given by the Berry-phase 
theory of DMI~\cite{mothedmisot,phase_space_berry,itsot,spicudmi}, 
which in the one-dimensional case runs
\bege\label{eq_dmi_berry}
D^{yx}\!\!=\!\!\int\!
\frac{d\, k_x}{2\pi}
\!\sum_{n}
\Bigl[
 f_{\vn{k}n}\mathcal{A}_{\vn{k}n}^{xx}
\!+\!\frac{1}{\beta}
\ln
\left[
1+
e^{-\beta(
\mathcal{E}_{\vn{k}n}
-\mu
)}
\right]
\mathcal{B}_{\vn{k}n}^{xx}
\Bigr].
\ee

\subsection{Two-dimensional electron gas without SOI}
\label{sec_gauge_field_rashba2d}
Due to Eq.~\eqref{eq_gauge_trafo_sigmax} the $U$ transformation
does not lead to simplifications in the case of the two-dimensional Rashba model
Eq.~\eqref{eq_rashba_model} when $\alpha\ne 0$. However, in the
case of $\alpha=0$ the $U$ transformation leads to a simplification of
Eq.~\eqref{eq_rashba_model}:
\bege
\begin{aligned}
\tilde{H}=&[U(x)]^{\dagger}
H
U(x)
=\\
=&-\frac{\hbar^2}{2m_e}
\left[
\Delta
-iq\sigma_y\frac{\partial}{\partial x}
-\frac{q^2}{4}
\right]+
\frac{\Delta V}{2}
\sigma_{z},
\end{aligned}
\ee
with corresponding crystal momentum representation
\bege\label{eq_rash_2d_tilde_k}
\begin{aligned}
&\tilde{H}_{\vn{k}}
=
\frac{\hbar^2}{2m_e}
\left[
\vn{k}^2
-q
k_{x}
\sigma_y
+\frac{q^2}{4}
\right]+
\frac{\Delta V}{2}
\sigma_{z}.
\end{aligned}
\ee
Since $\tilde{H}_{\vn{k}}$ is position-independent, its eigenvalues $\mathcal{E}_{\vn{k},q,n}$
can be determined easily. The free energy density is then obtained from
\bege
F_{q}=
-\frac{1}{\beta}
\int\frac{d^2 k}{(2\pi)^2}\sum_{n}
\ln
\left[
1+e^{-\beta(\mathcal{E}_{\vn{k},q,n}-\mu)}
\right]
\ee
and Eq.~\eqref{eq_gauge_approach_exchange} can be used 
to determine the exchange constant $\mathscr{A}^{xx}$.
\section{Exchange constants in model systems}
\label{sec_num_results}
\subsection{One-dimensional Rashba model}
\label{sec_one_dim_rashba}
\begin{figure}
\includegraphics[width=\linewidth]{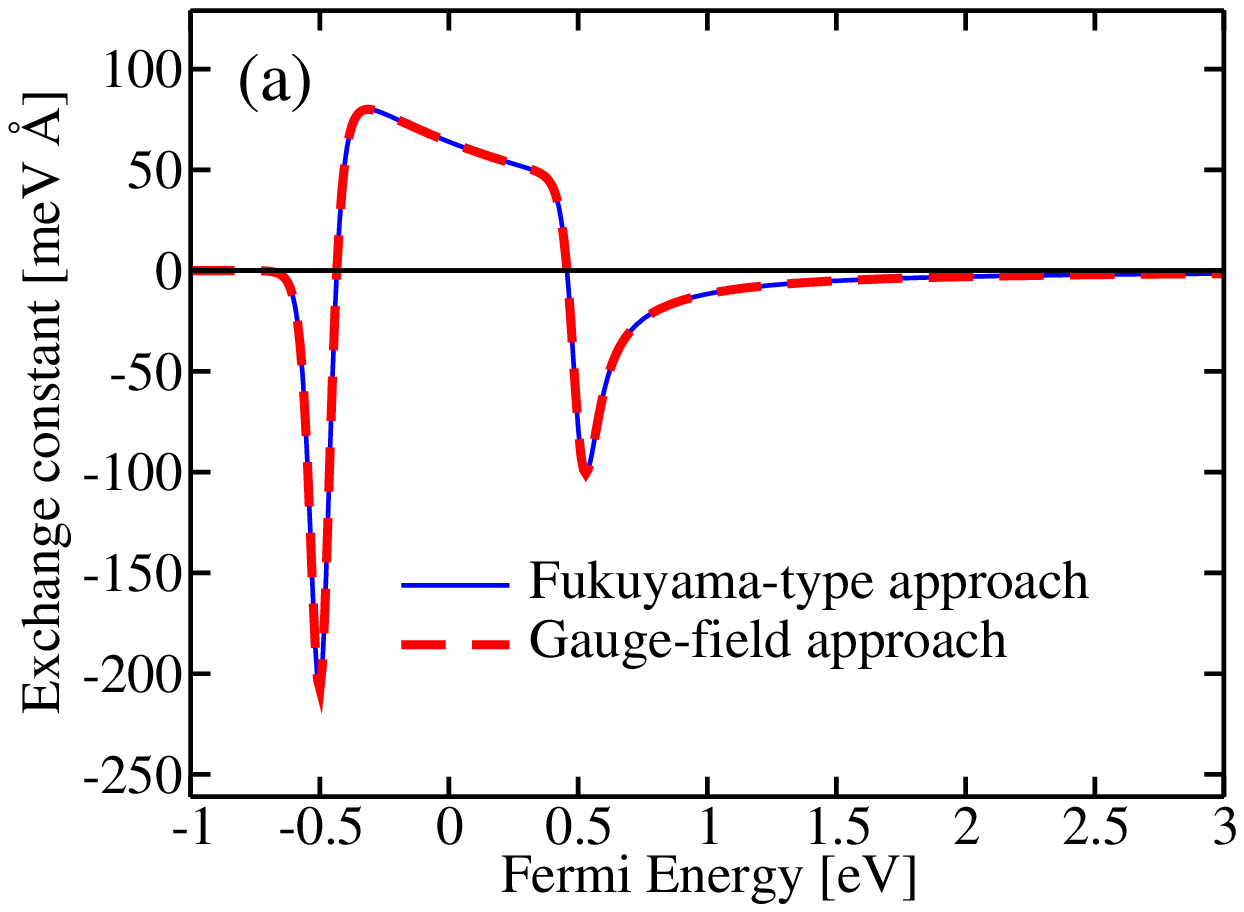}
\includegraphics[width=\linewidth]{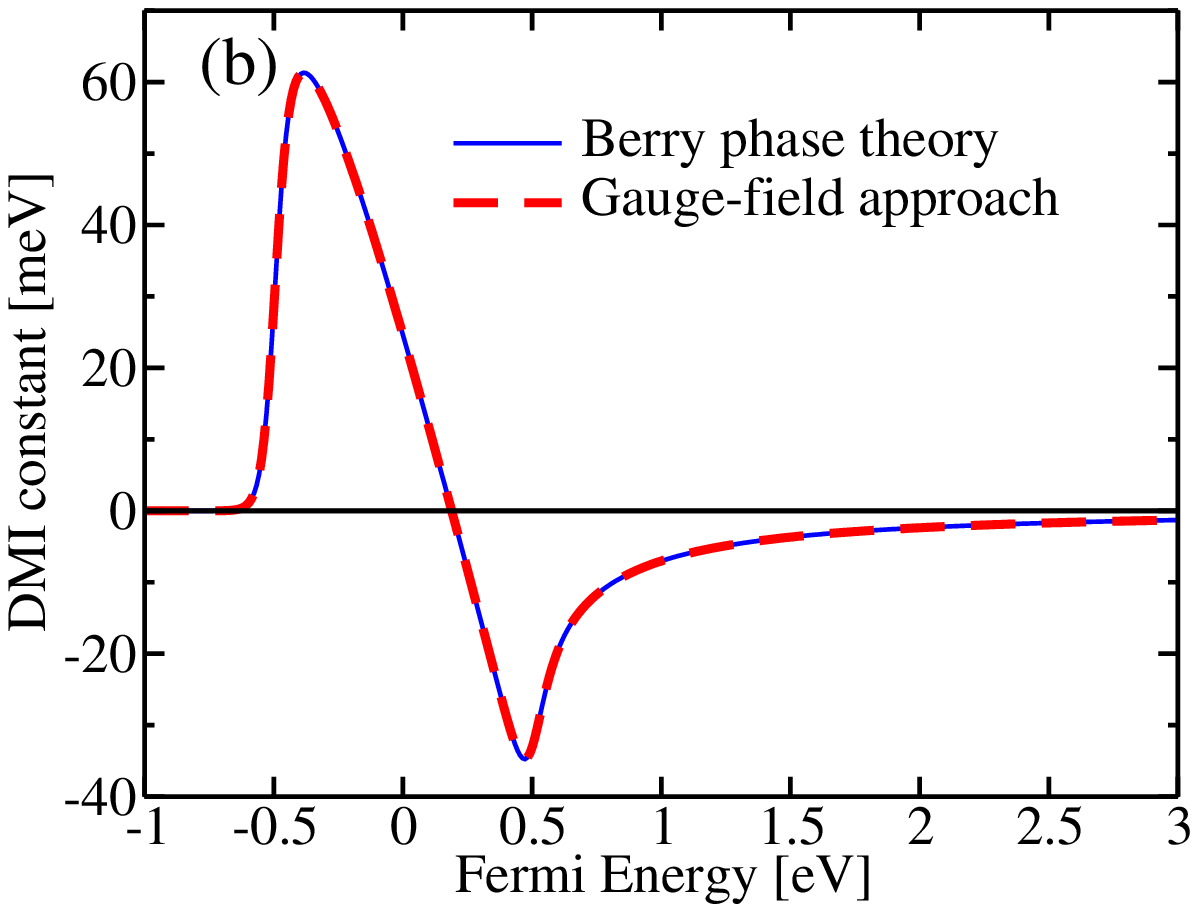}
\includegraphics[width=\linewidth]{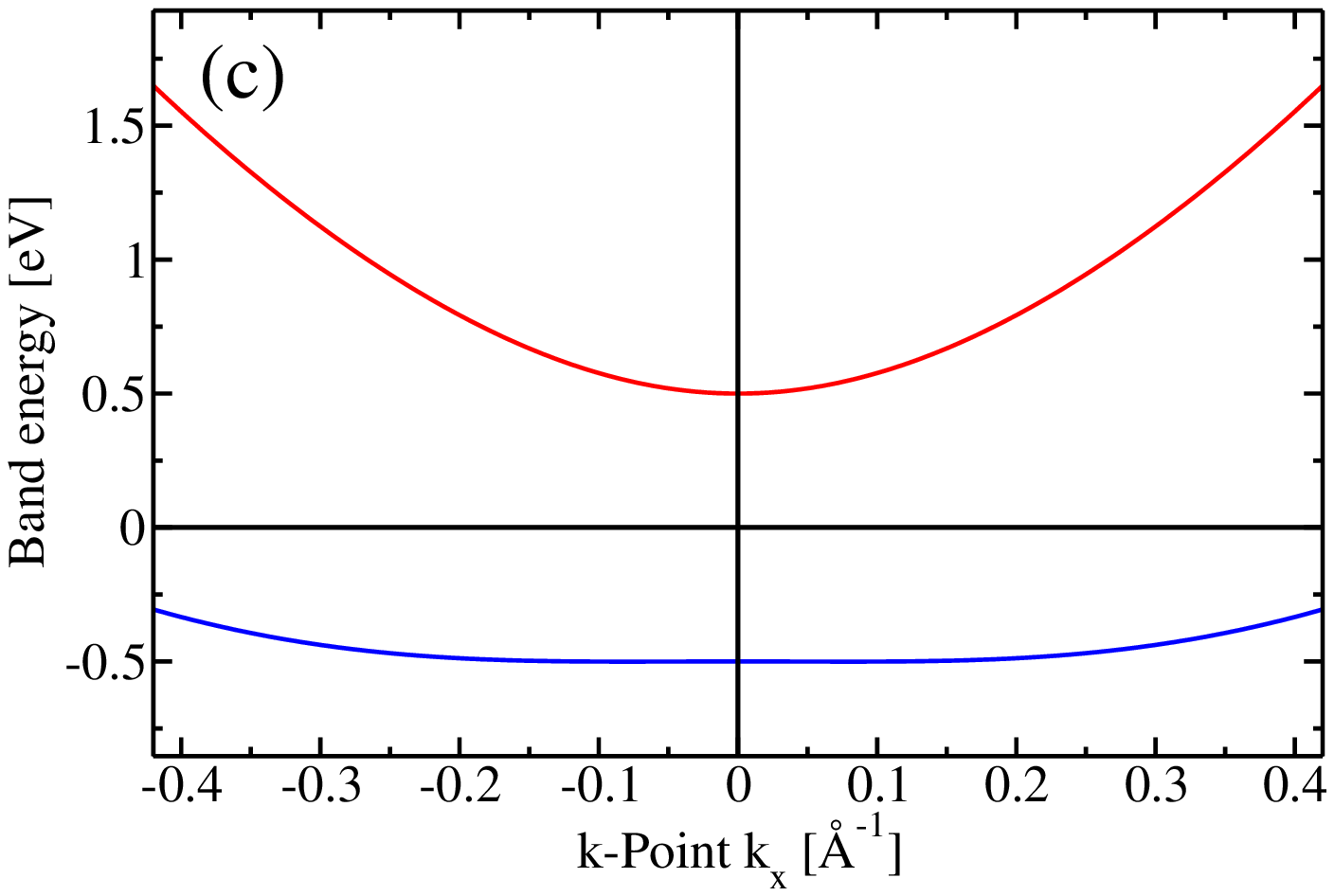}
\caption{\label{fig_gauge_vs_fukuyama}
(a) Exchange constant $\mathscr{A}^{xx}$ and
(b) DMI constant $D^{yx}$
in the one-dimensional 
Rashba model Eq.~\eqref{eq_rashba_model_onedim}
as a function of Fermi energy for the
model parameters $\Delta V=1$eV and $\alpha=2$eV\AA{}. 
Results obtained from
the gauge-field approach (dashed lines)
agree respectively to the exchange constant from the
Fukuyama-type approach 
and to the DMI-constant from the Berry-phase
approach (solid lines). 
(c) Band structure of the one-dimensional Rashba model.
}
\end{figure}

\begin{figure}
\includegraphics[width=\linewidth]{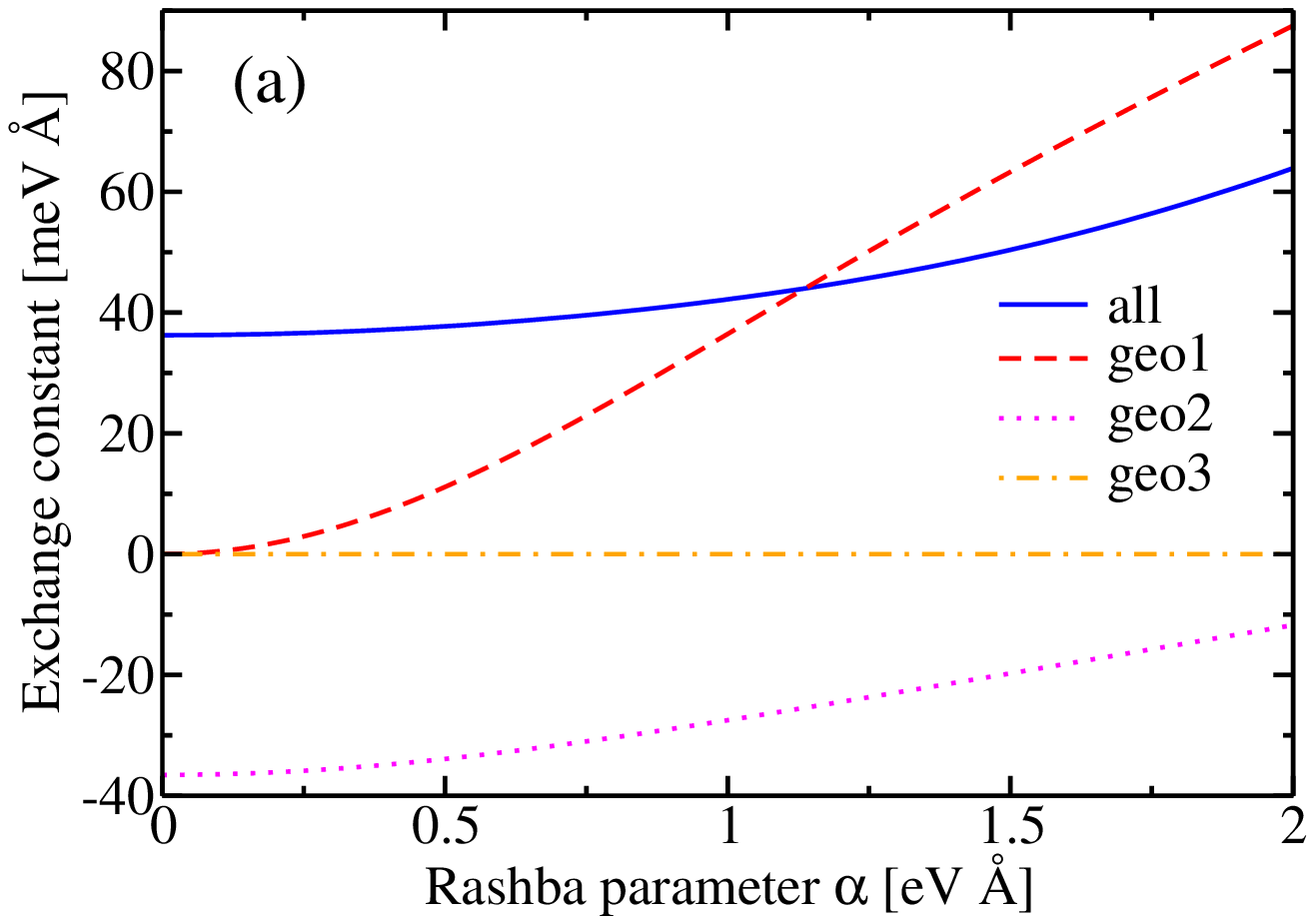}
\includegraphics[width=\linewidth]{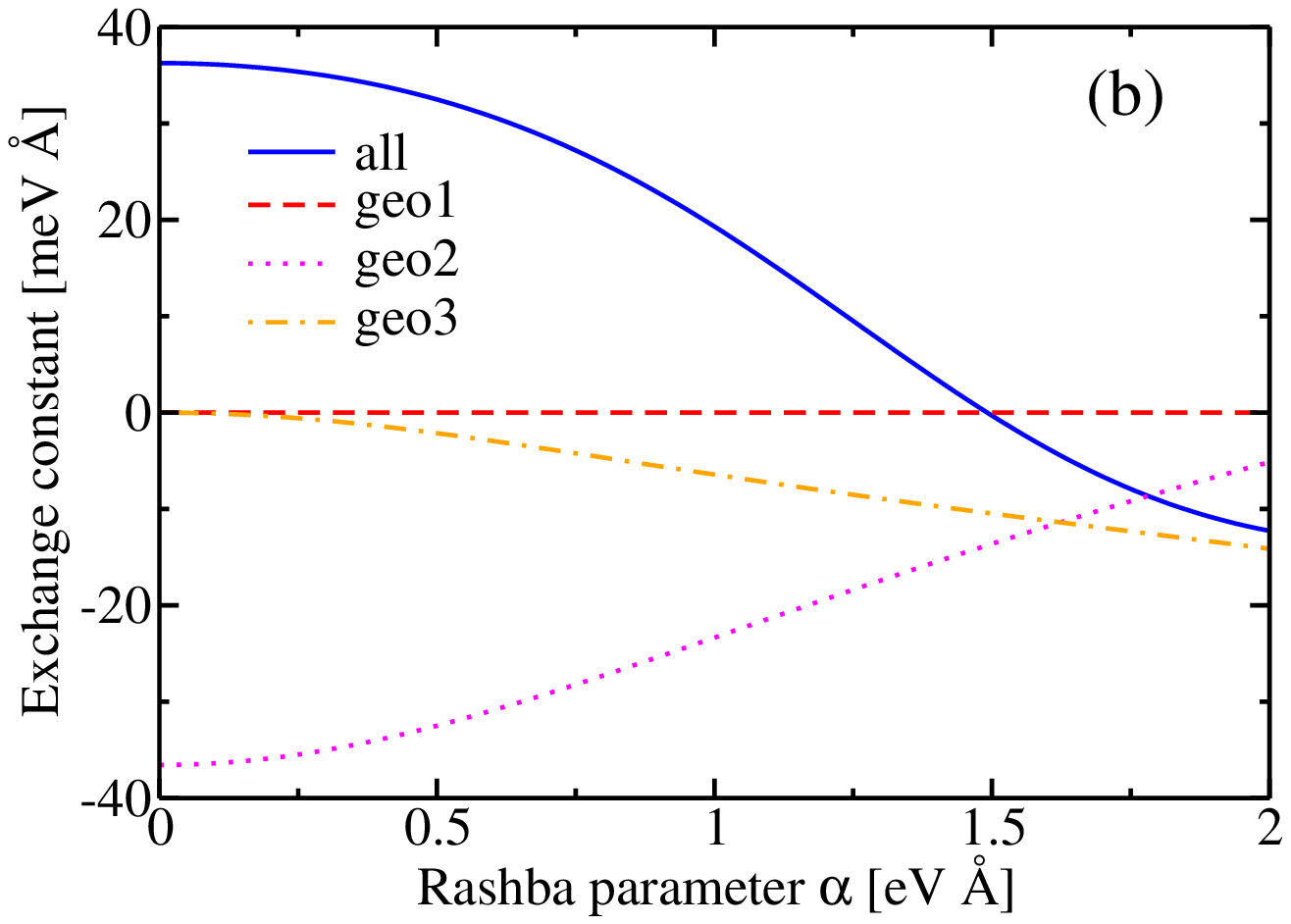}
\caption{\label{fig_axx_vs_axy}
(a) Exchange constant $\mathscr{A}^{xx}$ 
and (b) $\mathscr{A}^{xy}$ 
in the one-dimensional 
Rashba model Eq.~\eqref{eq_rashba_model_onedim}
as a function of the Rashba parameter $\alpha$. The
Fermi energy is set to zero and $\Delta V=1$eV. 
Solid lines: Complete exchange constants.
The geometrical contributions 
$\mathscr{A}_{\rm geo1}$ (dashed),
$\mathscr{A}_{\rm geo2}$ (dotted),
and
$\mathscr{A}_{\rm geo3}$ (dashed-dotted)
as defined in Eqs.~\eqref{eq_geo1} 
through \eqref{eq_geo3}
are shown as well.
}
\end{figure}

In the presence of SOI both the exchange constant $\mathscr{A}^{ij}$ as
obtained from Eq.~\eqref{eq_exchange_params_fukuyama} and the
DMI constant $D^{ij}$ as obtained from Eq.~\eqref{eq_dmi_berry} may
depend on the magnetization direction $\hat{\vn{n}}$. However, as
we explained in the discussion below Eq.~\eqref{eq_rashba_model_onedim}, 
rotations in spin-space around the $y$ axis are a symmetry operation
of the one-dimensional Rashba model. Since we consider the
special case of a cycloidal 
spin-spiral, Eq.~\eqref{eq_spin_spiral_cycloid}, which describes
a magnetization that rotates around the $y$ axis as one moves 
along the spin-spiral, $\mathscr{A}^{ij}$ 
and $D^{ij}$ are constant along this spin-spiral. This allows us to
compare the values of $\mathscr{A}^{ij}$ and $D^{ij}$
obtained for $\hat{\vn{n}}$ in $z$ direction to the values obtained
from the gauge-field approach in this particular case, while in
a general case a spin-spiral calculation will correspond to an 
$\hat{\vn{n}}$-integration 
of $\hat{\vn{n}}$-dependent $\mathscr{A}^{ij}$ and $D^{ij}$.

In Fig.~\ref{fig_gauge_vs_fukuyama} we show the
exchange constant $\mathscr{A}^{xx}$ as well as the DMI coefficient
for the one-dimensional Rashba 
model, Eq.~\eqref{eq_rashba_model_onedim}, 
as a function of the Fermi energy.
The parameters used in the model 
are $\Delta V=1$eV and $\alpha=2$eV\AA{} 
and we
set the temperature in the Fermi 
functions to $k_{\rm B}T=25$~meV.
Two approaches are compared: The dashed lines show the
results obtained from Eq.~\eqref{eq_gauge_approach_exchange}
and Eq.~\eqref{eq_gauge_approach_dmi} within the gauge-field approach,
where we used a small spin-spiral vector 
of $q=0.006$~\AA$^{-1}$ (we checked that making $q$ smaller does not 
affect the results). The solid line in Fig.~\ref{fig_gauge_vs_fukuyama}(a)
is obtained from the Fukuyama-type 
expression Eq.~\eqref{eq_exchange_params_fukuyama} 
for the exchange constant.
The solid line in  Fig.~\ref{fig_gauge_vs_fukuyama}(b)
is obtained from the Berry-phase theory of DMI, Eq.~\eqref{eq_dmi_berry}.
The results from the different methods are in perfect agreement,
which shows in particular 
that Eq.~\eqref{eq_exchange_params_fukuyama} can be used 
for calculating exchange constants even in the presence of SOI.
The exchange constant becomes negative 
when the Fermi energy is close to $\pm 0.5$~eV, i.e., close to
the band minima (see Fig.~\ref{fig_gauge_vs_fukuyama}(c)).   
Negative exchange constants imply that the ferromagnetic state is
unstable and that a spin-spiral state will form.
With increasing Fermi energy, the effect of Rashba SOI on the Fermi
surface becomes smaller and smaller. At very high Fermi energy the
Fermi surfaces with an without SOI differ very little. As a
consequence, the DMI is suppressed at high Fermi energy. 

We have verified that the gauge-field
approach, Eq.~\eqref{eq_gauge_approach_dmi}, 
and the Berry-phase theory, Eq.~\eqref{eq_dmi_berry}, agree at all orders in SOI.
Previously, we have shown~\cite{spicudmi} that the Berry-phase theory
reduces to the ground-state 
spin current~\cite{dmi_doppler_shift} 
at the first order in SOI. 

In Fig.~\ref{fig_axx_vs_axy} we show the exchange
constants $\mathscr{A}^{xx}$ and $\mathscr{A}^{xy}$
as a function of the Rashba parameter $\alpha$ when the
Fermi energy is set to zero and $\Delta V=1$eV.
$\mathscr{A}^{xx}$ is the exchange constant of a cycloidal spin-spiral
and $\mathscr{A}^{xy}$ is the one of a helical spin-spiral.
In the absence of SOI rotations in spin-space leave
the spectrum of the Hamiltonian invariant and therefore $\mathscr{A}^{xx}=\mathscr{A}^{xy}$.
For $\alpha\ne0$ $\mathscr{A}^{xx}$ and $\mathscr{A}^{xy}$ differ from
each other and
the difference becomes large with increasing $\alpha$.
The three geometrical contributions as defined  in Eqs.~\eqref{eq_geo1} 
through \eqref{eq_geo3} are shown in Fig.~\ref{fig_axx_vs_axy}
as well. 
The mixed Berry curvature and the mixed quantum metric
are zero without SOI and therefore we expect that
$\mathscr{A}_{\rm geo1}$ and $\mathscr{A}_{\rm geo3}$, which depend on
the mixed Berry curvature and the mixed quantum metric, differ
between cycloidal and helical spin-spirals, which is indeed the case: 
While $\mathscr{A}^{xx}_{\rm geo1}$
increases strongly with $\alpha$,
$\mathscr{A}^{xy}_{\rm geo1}$ is zero
and while $\mathscr{A}^{xx}_{\rm geo3}$ is zero,
$\mathscr{A}^{xy}_{\rm geo3}$ becomes negative with 
increasing $\alpha$.
In contrast, $\mathscr{A}^{xx}_{\rm geo2}$ and $\mathscr{A}^{xy}_{\rm geo2}$
are very similar, because they only involve the quantum metric in real
space as well as the inverse effective mass in $k$-space. Generally, the
geometrical contribution 
cannot be neglected and
is of the same order of 
magnitude as the total exchange constant.

\begin{figure}
\includegraphics[width=\linewidth]{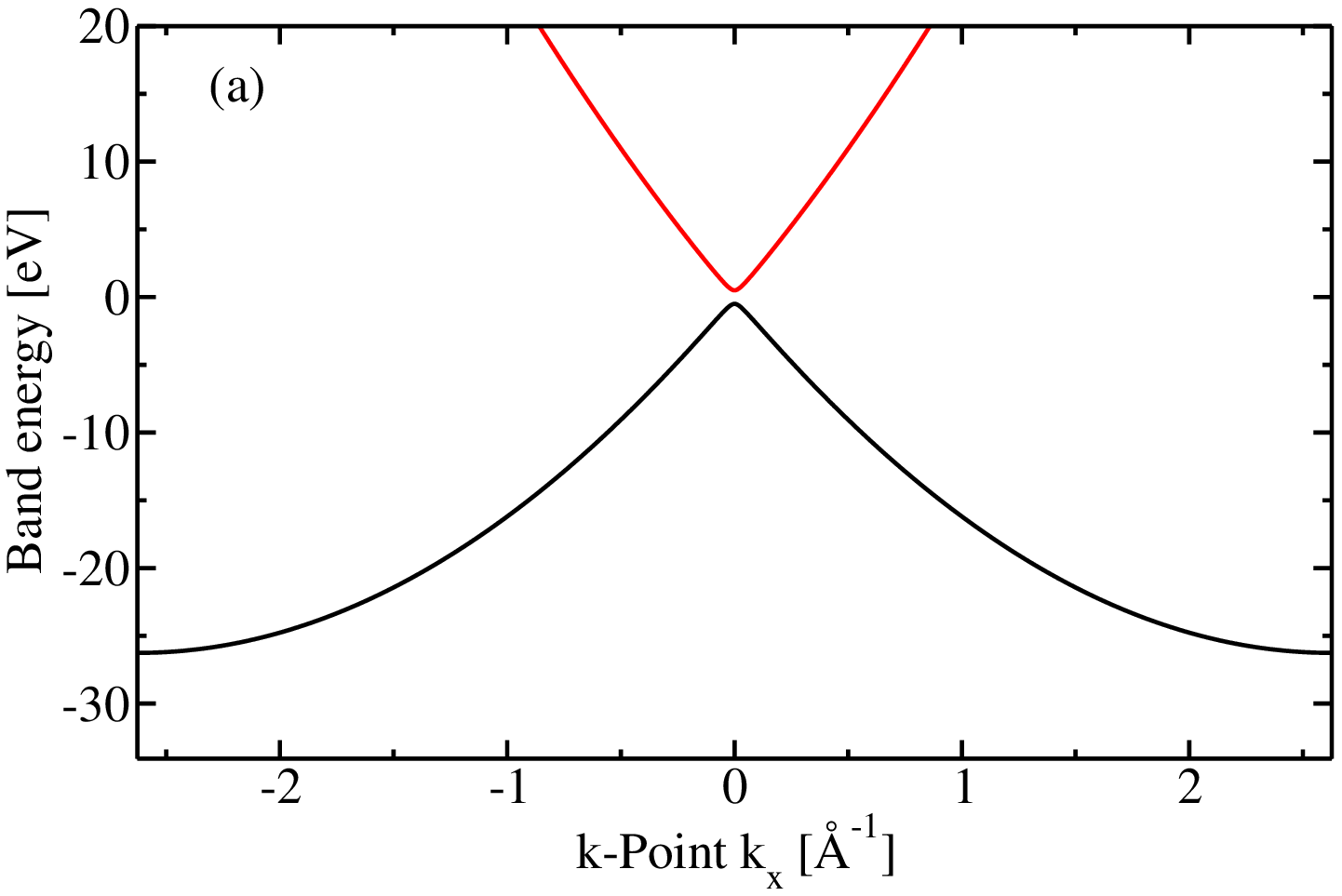}
\includegraphics[width=\linewidth]{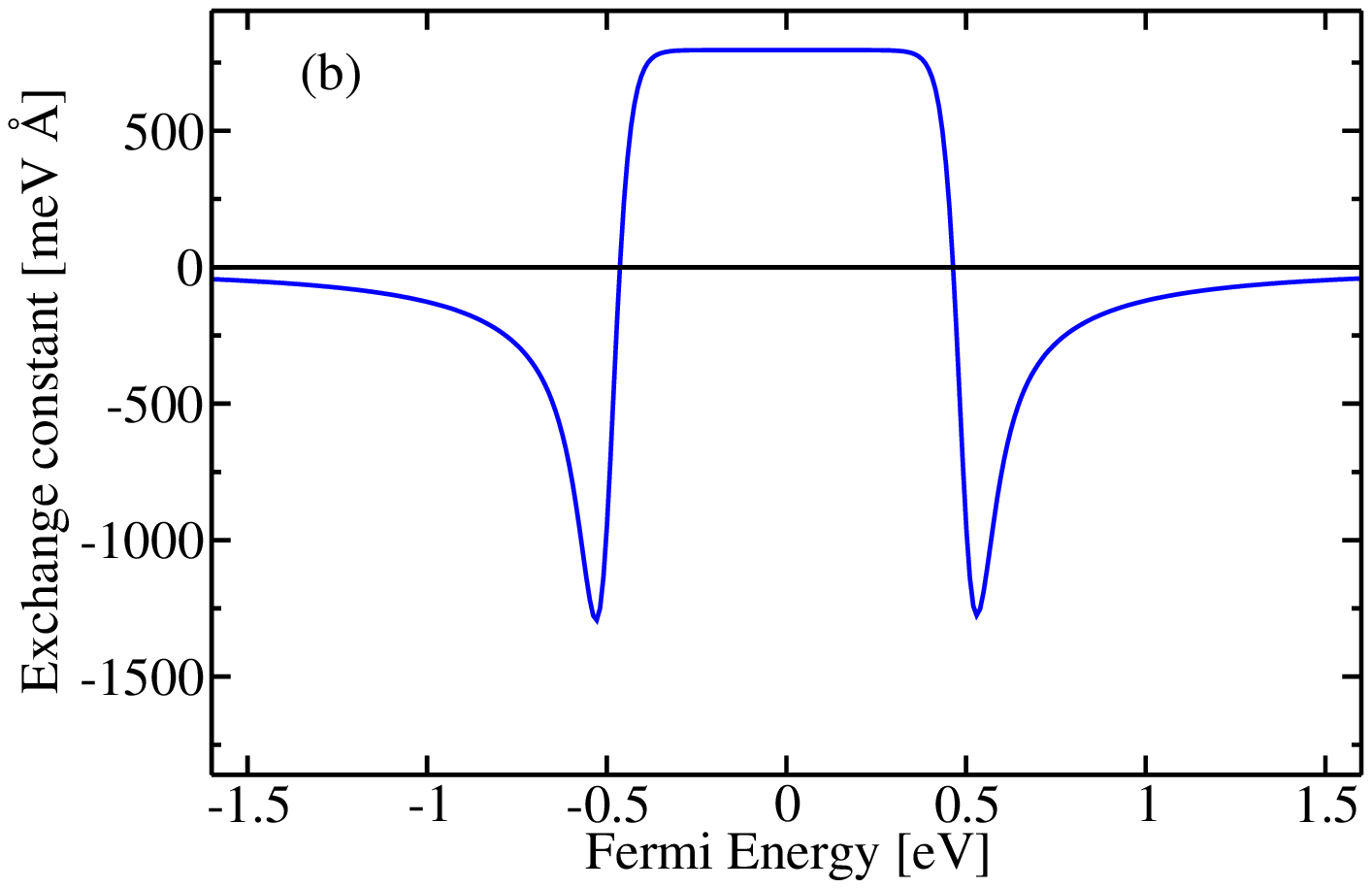}
\caption{\label{fig_1d_rashba_insulator}
(a) Band energy 
of the one-dimensional 
Rashba model Eq.~\eqref{eq_rashba_model_onedim}
with model parameters $\alpha=20$eV\AA{} 
and $\Delta V=1$eV. 
(b) 
The corresponding exchange constant exhibits a plateau 
between -0.3eV and 0.3eV due to the gap of the band structure in (a). 
}
\end{figure}

The expressions Eq.~\eqref{eq_exchange_params_semicla},
Eq.~\eqref{eq_axx_pol},
and Eq.~\eqref{eq_axx_inter}
contain both Fermi surface and Fermi sea terms.
The exchange constant does not vanish in band insulators
due to the Fermi sea terms and exhibits a plateau in the gap.
To illustrate this we show in Fig.~\ref{fig_1d_rashba_insulator}
the exchange constant of the one-dimensional Rashba model
with model parameters $\alpha=20$eV\AA{} 
and $\Delta V=1$eV. In the $k_x$ integration we use a 
cutoff of 2.63\AA{}$^{-1}$. This cutoff is necessary in 
order to obtain an insulating system because there is no global gap in 
the bandstructure of the
one-dimensional Rashba model. However, when we restrict the
range of $k$ points to the 
region -2.63\AA{}$^{-1}<k_x<$2.63\AA{}$^{-1}$ the 
band structure appears gapped as shown in
Fig.~\ref{fig_1d_rashba_insulator}(a).
As shown in Fig.~\ref{fig_1d_rashba_insulator}(b) the
corresponding exchange constant exhibits a plateau in the gap.

\subsection{Rashba model in two dimensions}
\label{sec_two_dim_rashba}
\begin{figure}
\includegraphics[width=\linewidth]{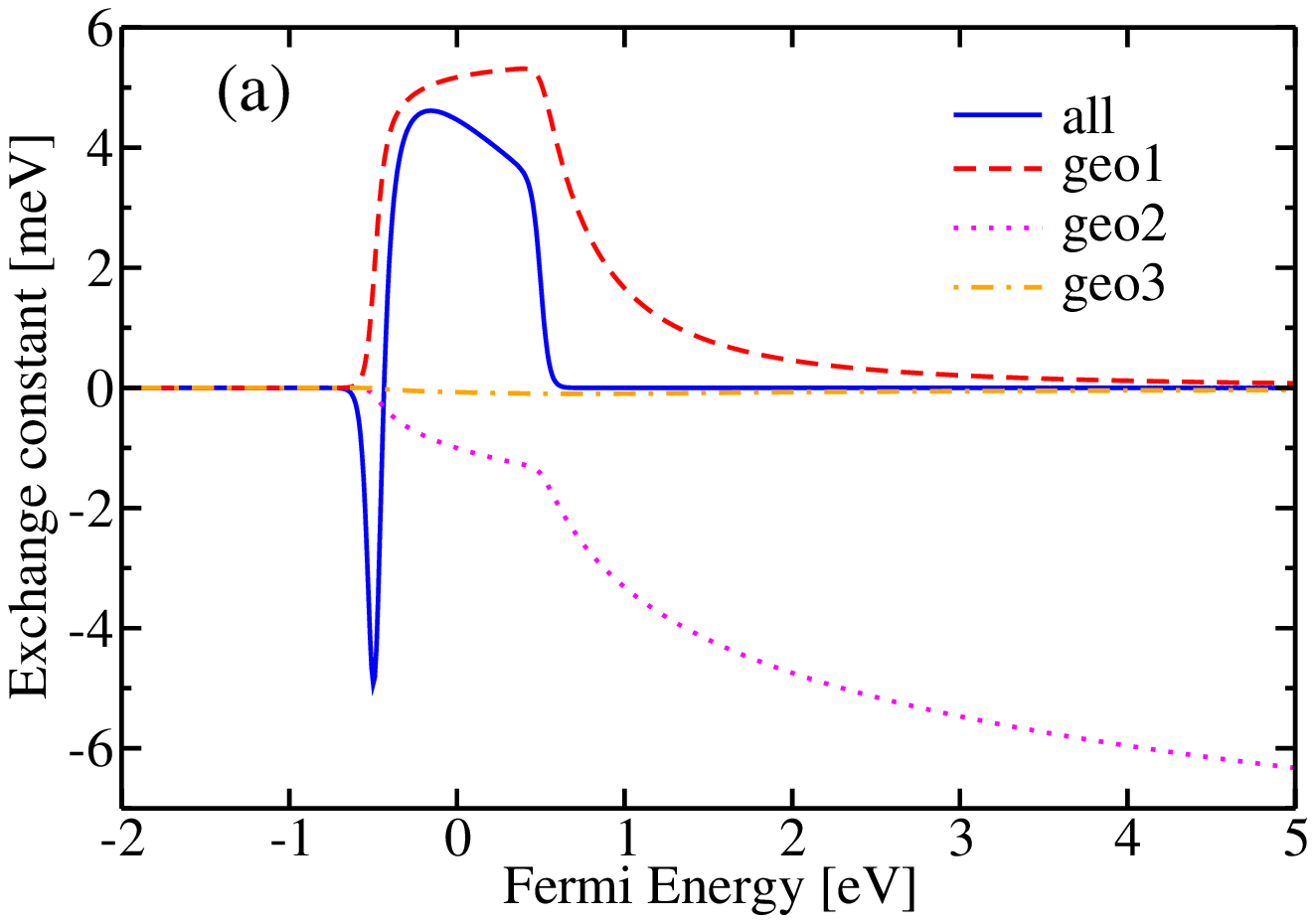}
\includegraphics[width=\linewidth]{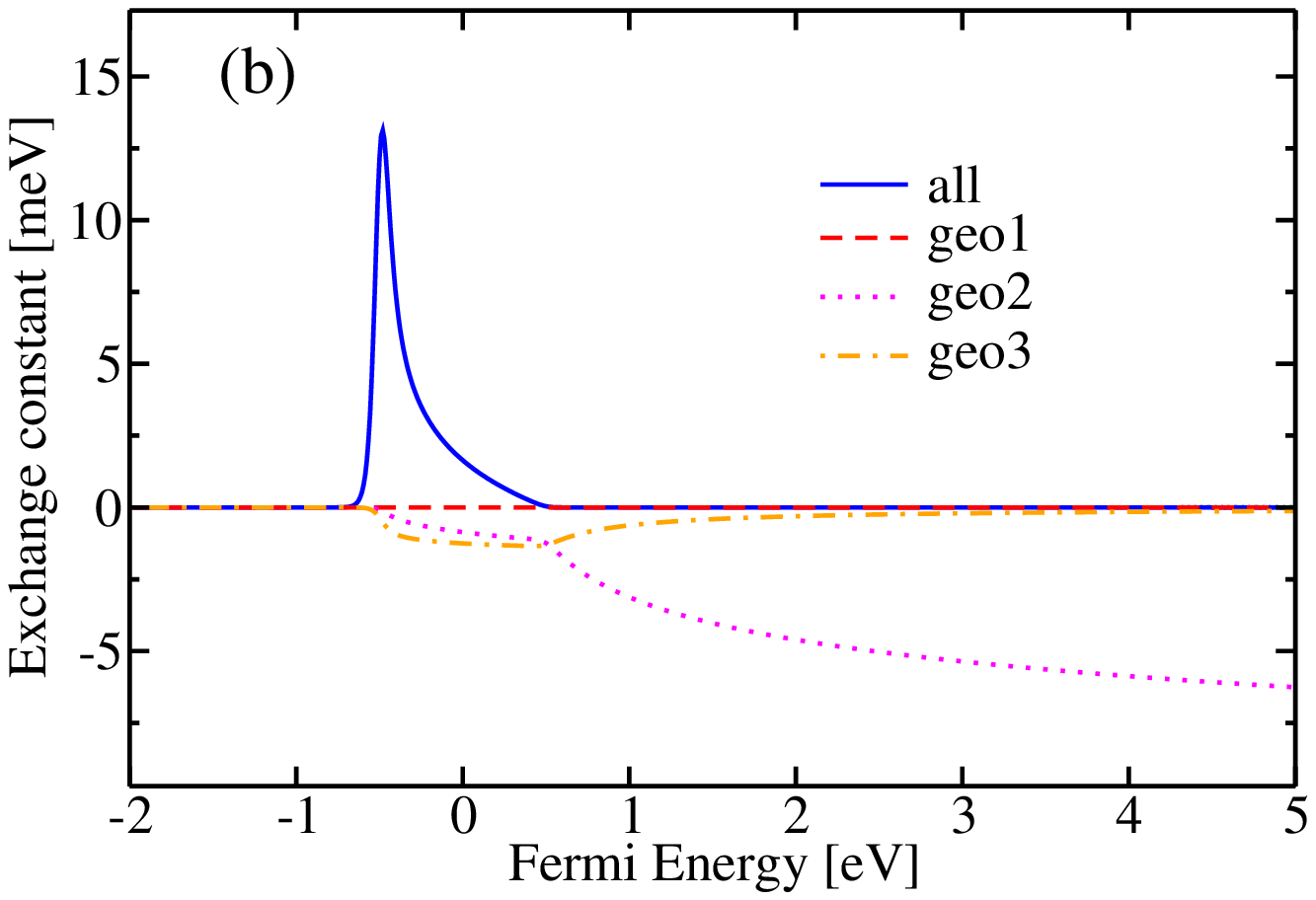}
\caption{\label{fig_2d_rashba}
(a) Exchange constant $\mathscr{A}^{xx}$ 
and (b) $\mathscr{A}^{xy}$ 
in the two-dimensional 
Rashba model Eq.~\eqref{eq_rashba_model}
as a function of the Fermi energy. The model parameters
are $\alpha=2$eV\AA{} 
and $\Delta V=1$eV. 
Solid lines: Complete exchange constants.
The geometrical contributions 
$\mathscr{A}_{\rm geo1}$ (dashed),
$\mathscr{A}_{\rm geo2}$ (dotted),
and
$\mathscr{A}_{\rm geo3}$ (dashed-dotted)
as defined in Eqs.~\eqref{eq_geo1} 
through \eqref{eq_geo3}
are shown as well.
}
\end{figure}
When the Rashba parameter $\alpha$ is zero 
the exchange constant of the two-dimensional Rashba model
can be obtained from the gauge-field approach as 
discussed in section~\ref{sec_gauge_field_rashba2d}. 
We checked that the gauge-field approach 
and Eq.~\eqref{eq_exchange_params_fukuyama} yield
identical results in this case.

We now turn to the case with $\alpha>0$, where we
use the model parameters $\alpha=2$eV\AA{} 
and $\Delta V=1$eV. 
In Fig.~\ref{fig_2d_rashba} we show the
exchange constants $\mathscr{A}^{xx}$
and $\mathscr{A}^{xy}$ as obtained 
from Eq.~\eqref{eq_exchange_params_fukuyama}
as a function of the Fermi energy, as well as the
geometrical contributions 
$\mathscr{A}_{\rm geo1}$,
$\mathscr{A}_{\rm geo2}$,
and
$\mathscr{A}_{\rm geo3}$
as defined in Eqs.~\eqref{eq_geo1} 
through \eqref{eq_geo3}. 
We rediscover several properties that we discussed already
in the one-dimensional Rashba model:
The exchange constant of the 
cycloidal spin-spiral ($\mathscr{A}^{xx}$) differs considerably
from the exchange constant of the helical spin-spiral ($\mathscr{A}^{xy}$)
when SOI is large. The contribution $\mathscr{A}_{\rm geo2}$ does not
differ much between helical spin-spiral and cycloidal spin-spiral, while
$\mathscr{A}_{\rm geo1}$ and $\mathscr{A}_{\rm geo3}$ are very different between
these two cases.
\section{Summary}
\label{sec_summary}
We derive a formula that expresses the exchange constants 
in terms
of Green's functions, velocity operators, and torque operators 
of a collinear ferromagnet. Thus, it allows us to access the
exchange constants directly from the electronic structure information
without the need for spin-spiral calculations. 
We compare this formula to Fukuyama's result for the orbital
magnetic susceptibility and find strong formal similarities between
these two theories. We rewrite the Green's function expression
for the exchange constant in terms of Berry curvatures and
quantum metrices in mixed phase-space. Thereby we identify
several geometrical contributions to the exchange constants
that we find to be generally important in free electron model systems.
Our formalism can
be used even in the presence of spin-orbit interaction, where
we find sizable differences between the exchange constants of 
helical and cycloidal spin spirals in the Rashba model.

\appendix
\section{From torque-operator expressions to curvatures and geometrical quantities}
\label{sec_appendix_torque}
In this appendix we discuss how to express matrix elements of the torque operator in terms
of derivatives with respect to the magnetization direction.
For this purpose we use that the Hamiltonian 
\bege\label{eq_app_hamil_coll_ferro}
\begin{aligned}
H(\vn{r})=&-\frac{\hbar^2}{2m_e}\Delta+V(\vn{r})+
\mubo
\vn{\sigma}\cdot \hat{\vn{n}}
\Bxc^{\rm xc}(\vn{r})+\\
&+
\frac{1}{2 e c^2}\mubo
\vn{\sigma}\cdot
\left[
\vn{\nabla}V(\vn{r})\times\vn{v}
\right].
\end{aligned}
\ee
is dependent on the
magnetization direction $\hat{\vn{n}}$ through the exchange interaction
$\mu_{\rm B}\vht{\sigma}\cdot\hat{\vn{n}}\,\Bxc^{\rm xc}(\vn{r})$. The derivative 
of $H$ with respect to magnetization direction $\hat{\vn{n}}$ 
can be expressed in terms of the torque operator:
\bege
\hat{\vn{n}}\times \frac{\partial H}{\partial \hat{\vn{n}}}=
\mu_{\rm B}\hat{\vn{n}}\times\vht{\sigma}\,\Bxc^{\rm xc}(\vn{r})=
-\mu_{\rm B}\vht{\sigma}\times\vn{\Bxc}^{\rm xc}(\vn{r})=
\vht{\mathcal{T}}(\vn{r}).
\ee
Thus, when the magnetization points in $z$ direction, i.e., 
when $\hat{\vn{n}}=\hat{\vn{e}}^z$, the
cartesian components of $\vht{\mathcal{T}}$
are given by
\bege
\begin{aligned}
\mathcal{T}^{x}&=-\frac{\partial H}{\partial \hat{n}^y  },
\\
\mathcal{T}^{y}&=\frac{\partial H}{\partial \hat{n}^x  }.
\\
\end{aligned}
\ee 
Using
\bege\label{eq_perturbation_in_k}
\begin{aligned}
\frac{\partial|u_{\vn{k}n}\rangle}{\partial k^x}
&=
\sum_{m\neq n}
\frac{|u_{\vn{k}m}\rangle\langle u_{\vn{k}m}|\frac{\partial H(\vn{k})}{\partial k^{x}}|u_{\vn{k}n}\rangle}
{\mathcal{E}^{\phantom{R}}_{\vn{k}n}-\mathcal{E}^{\phantom{R}}_{\vn{k}m}}\\
&+ia_{\vn{k}n}|u_{\vn{k}n}\rangle\\
&=
\hbar\sum_{m\neq n}
\frac{|u_{\vn{k}m}\rangle\langle u_{\vn{k}m}|v^{x}(\vn{k})|u_{\vn{k}n}\rangle}
{\mathcal{E}^{\phantom{R}}_{\vn{k}n}-\mathcal{E}^{\phantom{R}}_{\vn{k}m}}\\
&+ia_{\vn{k}n}|u_{\vn{k}n}\rangle
\end{aligned}
\ee
and
\bege\label{eq_perturbation_theory_for_angles}
\begin{aligned}
\frac{\partial|u_{\vn{k}n}\rangle}{\partial \hat{n}^x}
=&
\sum_{m\neq n}
\frac{|u_{\vn{k}m}\rangle\langle u_{\vn{k}m}|
\frac{\partial H(\vn{k})}{\partial \hat{n}^x }
|u_{\vn{k}n}\rangle}
{\mathcal{E}^{\phantom{R}}_{\vn{k}n}-\mathcal{E}^{\phantom{R}}_{\vn{k}m}}\\
&+i b_{\vn{k}n}|u_{\vn{k}n}\rangle\\
&=
\sum_{m\neq n}
\frac{|u_{\vn{k}m}\rangle\langle u_{\vn{k}m}|\mathcal{T}^{y}|u_{\vn{k}n}\rangle}
{\mathcal{E}^{\phantom{R}}_{\vn{k}n}-\mathcal{E}^{\phantom{R}}_{\vn{k}m}}\\
&+i b_{\vn{k}n}|u_{\vn{k}n}\rangle,
\end{aligned}
\ee
where the phases $a_{\vn{k}n}$ and $b_{\vn{k}n}$ determine
the gauge, $H(\vn{k})=e^{-i\vn{k}\cdot\vn{r}}He^{i\vn{k}\cdot\vn{r}}$ 
is the Hamiltonian
in crystal momentum representation 
and $u_{\vn{k}n}(\vn{r})=e^{-i\vn{k}\cdot\vn{r}}\psi_{\vn{k}n}(\vn{r})$ is
the lattice periodic part of the Bloch function $\psi_{\vn{k}n}(\vn{r})$, 
we can express
the
 mixed Berry curvature
in terms of the torque operator and the velocity operator
as follows:
\bege\label{eq_app_mixed_curvature}
\begin{aligned}
\mathcal{B}^{xx}_{\vn{k}n}=&
-2\,
{\rm Im}
\left\langle
\frac{\partial u_{\vn{k}n}}{ \partial\hat{n}^{x} }
\left|
\frac{\partial u_{\vn{k}n}}{\partial k^{x}}\right.
\right\rangle\\
=&
-2\hbar\,
{\rm Im}
\sum_{m\neq n}
\frac{
\langle \psi_{\vn{k}n}  |\mathcal{T}^{y}| \psi_{\vn{k}m}  \rangle
\langle \psi_{\vn{k}m}  |v^{x}| \psi_{\vn{k}n}  \rangle
}
{(\mathcal{E}^{\phantom{R}}_{\vn{k}m}-
\mathcal{E}^{\phantom{R}}_{\vn{k}n})^{2}}.
\end{aligned}
\ee
Similarly, we obtain an expression for the
mixed quantum metric in terms of the torque
operator and the velocity operator:
\bege\label{eq_app_mixed_quantum_metric}
\begin{aligned}
\mathcal{G}^{xx}_{\vn{k}n}&={\rm Re}
\left[
\frac{
\partial
\langle u_{\vn{k}n}|
}
{\partial \hat{n}^x}
\Bigl[
1-
|u_{\vn{k}n}\rangle
\langle u_{\vn{k}n}|
\Bigr]
\frac{\partial
|u_{\vn{k}n}\rangle
}
{\partial k^x}
\right]\\
&=
\hbar\,
{\rm Re}
\sum_{m\neq n}
\frac{
\langle \psi_{\vn{k}n}  |\mathcal{T}^{y}| \psi_{\vn{k}m}  \rangle
\langle \psi_{\vn{k}m}  |v^{x}| \psi_{\vn{k}n}  \rangle
}
{(\mathcal{E}^{\phantom{R}}_{\vn{k}m}-
\mathcal{E}^{\phantom{R}}_{\vn{k}n})^{2}}.
\end{aligned}
\ee
The
quantum metric in magnetization space
can be written as
\bege\label{eq_app_quantum_metric_mag}
\begin{aligned}
\tilde{g}^{xx}_{\vn{k}n}&={\rm Re}
\left[
\frac{
\partial
\langle u_{\vn{k}n}|
}{\partial \hat{n}^x}
\Bigl[
1-
|u_{\vn{k}n}\rangle
\langle u_{\vn{k}n}|
\Bigr]
\frac{\partial
|u_{\vn{k}n}\rangle
}{\partial \hat{n}^x}
\right]\\
&=
{\rm Re}
\sum_{m\neq n}
\frac{
\langle \psi_{\vn{k}n}  |\mathcal{T}^{y}| \psi_{\vn{k}m}  \rangle
\langle \psi_{\vn{k}m}  |\mathcal{T}^{y}| \psi_{\vn{k}n}  \rangle
}
{(\mathcal{E}^{\phantom{R}}_{\vn{k}m}-
\mathcal{E}^{\phantom{R}}_{\vn{k}n})^{2}}.
\end{aligned}
\ee
The
twist-torque moment of wavepackets is given by
\bege
\begin{aligned}
\mathcal{A}^{xx}_{\vn{k}n}&=
-
{\rm Im}
\left\langle
\frac{\partial u_{\vn{k}n}}{ \partial\hat{n}^{x} }
\right|
\!\Bigl[
\mathcal{E}_{\vn{k}n}-H_{\vn{k}}
\Bigr]
\!\left|
\frac{\partial u_{\vn{k}n}}{\partial k^{x}}
\right\rangle\\
&=
\hbar\,
{\rm Im}
\sum_{m\neq n}
\frac{
\langle \psi_{\vn{k}n}  |\mathcal{T}^{y}| \psi_{\vn{k}m}  \rangle
\langle \psi_{\vn{k}m}  |v^{x}| \psi_{\vn{k}n}  \rangle
}
{(\mathcal{E}^{\phantom{R}}_{\vn{k}m}-
\mathcal{E}^{\phantom{R}}_{\vn{k}n})}.
\end{aligned}
\ee
The interband Berry connection in magnetization space
can be written as
\bege
\begin{aligned}
\bar{A}^{x}_{\vn{k}mn}&=i
\langle u_{\vn{k}m}|
\frac{\partial
|u_{\vn{k}n}\rangle
}{\partial \hat{n}^{x}}=\\
&=i
\frac{\langle u_{\vn{k}m}|\mathcal{T}^{y}|u_{\vn{k}n}\rangle}
{\mathcal{E}^{\phantom{R}}_{\vn{k}n}-\mathcal{E}^{\phantom{R}}_{\vn{k}m}}.
\end{aligned}
\ee
The
mixed phase-space analogue of the inverse effective mass tensor can
be expressed in terms of the torque operator as
\bege
\begin{aligned}
\bar{\alpha}^{xx}_{\vn{k}n}&=
\frac{\partial^2 \mathcal{E}_{\vn{k}n}}{\partial k^x \partial\hat{n}^x}\\
&=
2\hbar\,
{\rm Re}
\sum_{m\neq n}
\frac{
\langle \psi_{\vn{k}n}  |\mathcal{T}^{y}| \psi_{\vn{k}m}  \rangle
\langle \psi_{\vn{k}m}  |v^{x}| \psi_{\vn{k}n}  \rangle
}
{(\mathcal{E}^{\phantom{R}}_{\vn{k}n}-
\mathcal{E}^{\phantom{R}}_{\vn{k}m})}.
\end{aligned}
\ee
\section{Analytical expressions when SOI is not included}
\label{sec_analytical}
In the following we derive analytical expressions for the case when 
SOI is not included in the Hamiltonian.
When SOI is not included in the Hamiltonian, one can show that
\bege
\mathcal{T}^{y}=\frac{i}{2}[H,\sigma^y].
\ee
Inserting this identity into Eq.~\eqref{eq_app_quantum_metric_mag}
one obtains the result
\bege
\tilde{g}^{xx}_{\vn{k}n}=\frac{1}{4}.
\ee
Inserting this result into Eq.~\eqref{eq_geo2} we get
\bege
\mathscr{A}_{\rm geo2}^{xx}=-\frac{1}{24}\frac{\hbar^2}{m_e}\mathcal{N},
\ee
where 
\bege
\mathcal{N}=\intkspa\sum_{n} f_{\vn{k}n} 
\ee
is the electron density. In the case $d=2$ we have
\bege
\mathcal{N}=\frac{m \mathcal{E}_{\rm F} }{\pi \hbar^2},
\ee
if both majority and minority bands are occupied.
If only the majority band is occupied we have instead
\bege
\mathcal{N}=\frac{m}{2\pi\hbar^2}
\left[
\mathcal{E}_{\rm F}+\frac{\Delta V}{2}
\right].
\ee
$\mathscr{A}_{\rm geo2}^{xx}$ is the second term 
in Eq.~\eqref{eq_exchange_params_semicla_nosoi}.
Similarly, the first term in
Eq.~\eqref{eq_exchange_params_semicla_nosoi} evaluates to
\bege
\intkspa
\sum_{n}
\frac{1}{3}f_{\vn{k}n}
\tilde{g}^{xx}_{\vn{k}n}
\frac{\hbar^2}{m_e}=\frac{1}{12}
\frac{\hbar^2}{m_e}\mathcal{N}.
\ee

The third term in Eq.~\eqref{eq_exchange_params_semicla_nosoi} 
can be written as
\bege\label{eq_app_term3}
\begin{aligned}
&-\frac{\hbar^2}{6}
\intkspa\sum_{n}
f'_{\vn{k}n}
v_{\vn{k}n}^{x}
\left\langle
\frac{\partial u_{\vn{k}n}}{ \partial  \hat{n}^{x} }
\right|
[
v^{x}_{\vn{k}}
+
v^{x}_{\vn{k}n}
]
\left|
\frac{\partial u_{\vn{k}n}}{\partial \hat{n}^{x}}
\right\rangle=\\
&=-\frac{\hbar^2}{12}
\intkspa\sum_{n}
f'_{\vn{k}n}
[v_{\vn{k}n}^{x}]^2.
\end{aligned}
\ee
When $d=2$ this becomes
\bege
\frac{\hbar^4}{48 \pi m^2}
\int
k^3 d k
\sum_{n}
\delta
\left(
\mathcal{E}_{\rm F}
-\frac{\hbar^2 k^2}{2m}
-\frac{\Delta V}{2}s_{\vn{k}n}
\right)
\ee
where $s_{\vn{k}n}=\pm 1$ denotes the spin (+1 for minority spin).
When only the majority band is occupied this is equal to
\bege
\frac{1}{24 \pi}
\left[
\mathcal{E}_{\rm F}+\frac{\Delta V}{2}
\right]
\ee
and when both minority and majority bands are occupied it is equal to
\bege
\frac{1}{12 \pi}
\mathcal{E}_{\rm F}.
\ee

The interband Berry connection in magnetization space
becomes:
\bege
\bar{A}^{x}_{\vn{k}mn}=
\frac{1}{2}
\langle
u _{\vn{k}n}
|
\sigma^{y}
|
u _{\vn{k}n}
\rangle.
\ee
Consequently,
the fourth term in Eq.~\eqref{eq_exchange_params_semicla_nosoi} 
can be written as
\bege
\begin{aligned}
&\frac{\hbar^2}{3}
\intkspa\sum_{n}
f_{\vn{k}n}
v^{x}_{\vn{k}n}
v^{x}_{\vn{k}n}
\sum_{m\ne n}
\frac{
\bar{A}^{x}_{\vn{k}mn}
[\bar{A}^{x}_{\vn{k}mn}]^{*}
}
{\mathcal{E}_{\vn{k}n}-\mathcal{E}_{\vn{k}m}}=\\
&=\frac{\hbar^2}{24 \Delta V}
\intkspa\sum_{n}
f_{\vn{k}n}
s _{\vn{k}n}
\frac{\hbar^2 k^2}{m^2}
=\\
&=-\frac{1}{24\pi}\mathcal{E}_{\rm F},
\end{aligned}
\ee
where the last line holds only in the case $d=2$ when both majority
and minority bands are occupied.
When only the majority band is occupied we obtain in the case $d=2$
for the fourth term
\bege
-\frac{1}{96\pi}
\left[
\mathcal{E}_{\rm F}+\frac{\Delta V}{2}
\right]
-\frac{1}{48\pi}
\frac{\mathcal{E}_{\rm F}}{\Delta V}
\left[
\mathcal{E}_{\rm F}+\frac{\Delta V}{2}
\right].
\ee

The fifth term in Eq.~\eqref{eq_exchange_params_semicla_nosoi} 
vanishes for the two-band model systems considered in this work.

For two-band models the sixth term in Eq.~\eqref{eq_exchange_params_semicla_nosoi} 
is simply twice the fourth term.

The seventh term in Eq.~\eqref{eq_exchange_params_semicla_nosoi} 
vanishes for the two-band model systems studied in this work.

Summing up all terms we obtain zero when both majority and minority
bands
are occupied. When only the majority band is occupied we obtain
\bege
\mathscr{A}^{xx}=\frac{1}{32\pi}
\left[
\mathcal{E}_{\rm F}+\frac{\Delta V}{2}
\right]
-
\frac{3}{48\pi}
\frac{\mathcal{E}_{\rm F}}{\Delta V}
\left[
\mathcal{E}_{\rm F}+\frac{\Delta V}{2}
\right]
\ee
in the case $d=2$.
\bibliography{geotexmosoi}

\end{document}